\documentclass[aps,prb,twocolumn,showpacs,superscriptaddress]{revtex4}
\usepackage{epsfig}
\usepackage{graphicx}
\usepackage{amsfonts}
\usepackage[figuresright]{rotating}
\usepackage{amssymb}
\usepackage{amsmath}
\def\avg#1{\langle#1\rangle}
\def\Re{\rm{Re}}
\def\Im{\rm{Im}}
\def\be{\begin{equation}} \def\ee{\end{equation}}
\def\bea{\begin{eqnarray}} \def\eea{\end{eqnarray}}

\def\nn{\nonumber}

\begin{document}
\title{Quantum magnetism of ultra-cold fermion systems with
the symplectic symmetry}
\author{Hsiang-Hsuan Hung}
\affiliation{Department of Physics, University of California, San
Diego, CA 92093}
\author{Yupeng Wang}
\affiliation{Beijing National Laboratory for Condensed Matter
Physics, Institute of Physics, Chinese Academy of Sciences, Beijing
100080, P. R. China}
\author{Congjun Wu}
\affiliation{Department of Physics, University of California, San
Diego, CA 92093}

\begin{abstract}
We numerically study quantum magnetism of ultra-cold alkali and
alkaline-earth fermion systems with large hyperfine spin
$F=\frac{3}{2}$, which are characterized by a generic $Sp(N)$
symmetry with $N=4$. The methods of exact diagonalization (ED) and
density-matrix-renormalization-group are employed for the large size
one-dimensional (1D) systems, and ED is applied to a two-dimensional
(2D) square lattice on small sizes. We focus on the magnetic
exchange models in the Mott-insulating state at quarter-filling.
Both 1D and 2D systems exhibit rich phase diagrams depending on the
ratio between the spin exchanges $J_0$ and $J_2$ in the bond spin
singlet and quintet channels, respectively. In 1D, the ground states
exhibit a long-range-ordered dimerization with a finite spin gap at
$J_0/J_2>1$, and a gapless spin liquid state at $J_0/J_2 \le 1$,
respectively. In the former and latter cases, the correlation
functions exhibit the two-site and four-site periodicities,
respectively. In 2D, various spin correlation functions are
calculated up to the size of $4\times 4$. The Neel-type spin
correlation dominates at large values of $J_0/J_2$, while a $2\times
2$ plaquette correlation is prominent at small values of this ratio.
Between them, a columnar spin-Peierls dimerization correlation
peaks. We infer the competitions among the plaquette ordering, the
dimer ordering, and the Neel ordering in the 2D system.
\end{abstract}
\pacs{71.10.Fd, 75.10.Jm, 71.10.Pm, 75.40.Mg}
\maketitle

\section{introduction}
\label{sect:intro}

The recent experimental progress on the ultracold
Fermi gases with large hyperfine spin provides an exciting
opportunity to investigate novel physical properties
\cite{Killian2010,takasu2010,wu2010, wu2006a}. In usual condensed
matter systems, large spin is not considered  particularly
interesting because large values of spin suppress quantum fluctuations.
For example, in transition metal oxides, a large spin on each cation
site is usually referred as an effective spin $S$ composed of $2S$
electrons by Hund's rule. The spin exchange between two cation sites
at the leading order of the perturbation theory only involves
swapping one pair of electrons regardless of how large $S$ is. The
variation of $S_z$ is only $\pm 1$, thus increasing $S$ reduces
quantum fluctuations known as the $1/S$-effect. In contrast,
in ultracold fermion systems, the situation is dramatically
different, in which large hyperfine spin enhances quantum
fluctuations. Each atom moves as a whole object carrying a large
hyperfine spin. Exchanging cold fermions can completely flip the
entire hyperfine-spin configuration, and thus enhances quantum
fluctuations. In other words, large spin physics in solid state
systems is usually in the large $S$-limit, while in cold atom
systems it is in the large $N$ limit where $N$ is the number of fermion
components $2F+1$ \cite{wu2010}.
We follow the convention in atomic physics to use $F$ to denote
the  hyperfine spin of the atom.

Ultracold fermion systems with large hyperfine spins have aroused a
great deal of theoretical interests. Early work studied the rich
structures of the Fermi liquid theory \cite{YIP1999} and the Cooper
pairing structures \cite{HO1999}. Considerable progress has been
made in the simplest large hyperfine spin systems with
$F=\frac{3}{2}$, whose possible candidate atoms are  $^{132}$Cs,
$^9$Be, $^{135}$Ba, $^{137}$Ba and $^{201}$Hg. These include both
alkaline-earth-like atoms with zero electron spin due to the fully
filled electron shells, and non-alkaline-earth atoms with nonzero
electron spins \cite{wu2006a,chen2005,xu2008}. In both cases, a
generic $Sp(4)$, or, isomorphically, $SO(5)$ symmetry is proved
without fine tuning. Such a high symmetry without fine-tuning is
rare in both condensed matter and cold atom systems. It brings
hidden degeneracy in the collective modes in the Fermi liquid theory
\cite{wu2007a}, fruitful patterns of quantum
magnetism\cite{WU2003,wu2006a,WU2005a,chen2005,xu2008} and Cooper
pairing with large internal spin degrees of freedom
\cite{wu_hu_zhang2010,xu2008}. Further investigations in the
community include the study of Mott insulating states
\cite{tu2006,tu2007,zheng2011,jiang2010,Lee2010,Lewenstein2010},
Beth-ansatz solution \cite{Controzzi2006, jiang2010a}, Kondo effect
\cite{hattori2005}, and the 4-fermion quartetting superfluidity
\cite{lecheminant2005,lecheminant2008,lee2007}. Recently, $SU(N)$
models have been proposed for the alkaline-earth fermion atoms since
their interactions are insensitive to their nuclear spins. It is a
special case of the $Sp(N)$ model by further tuning interaction
parameters of spin singlet and multiplet channels to be the same
\cite{gorshkov2010,hermele2009,xu2010}. The possible ferromagnetic
states have also been studied for the $SU(6)$ symmetric system of
$^{173}$Yb \cite{cazalilla2009}. A detailed summary is presented in
a review Ref. [\onlinecite{wu2006a}] and a non-technique
introduction is published at Ref. [\onlinecite{wu2010}] by one of
the authors. In a different context of heavy fermion systems, the
effects of sympletic symmetry to quantum magnetism have also been
studied in Ref. [\onlinecite{flint2008,flint2009}].

One dimensional (1D) systems are important for the study of strong
correlation physics because of the dominant interaction effects.
Furthermore, controllable analytical and numeric methods are
available. In Ref. [\onlinecite{WU2005a}], one of the authors
performed the bosonization method to study competing phases in 1D
systems with $F=\frac{3}{2}$, including the gapless Luttinger
liquid, spin gapped Luther-Emery liquid with Cooper pairing
instability, and 4-fermion quartetting superfluid at incommensurate
fillings. At commensurate fillings with strong repulsive
interactions, a charge gap opens and the systems become
Mott-insulating. The gapless Luttinger liquid phase becomes a
gapless spin liquid phase at quarter-filling and dimerized at
half-filling, respectively \cite{wu2006a}. The Luther-Emery phase
becomes the gapped $Sp(4)$ dimer phase at quarter-filling and the
on-site singlet phase at half-filling, respectively \cite{wu2006a}.

On the other hand, the two dimensional (2D) $Sp(4)$ Heisenberg model
is still far away from clear understanding. Such a system can bring
fruitful intriguing features of quantum magnetism which do not
exhibit in usual solid state systems. For example, in the special
case of the $SU(4)$ symmetry, four particles are required to form an
$SU(4)$ singlet, thus its quantum magnetism is characterized by the
4-site correlation beyond two sites. Such a state is the analogy to
the three-quark color singlet baryon state in quantum
chromodynamics. It is also the magnetism counterpart of the
4-fermion quartetting instability with attractive interactions
\cite{wu2006a}. Recently a magnetic phase diagram in a spatially
anisotropic square lattice of the $Sp(4)$ quantum magnetism is
provided by means of large-$N$ field-theoretical
approach\cite{kolezhuk2010}. A phase transition between the
long-range Neel order state and the disordered valence bond solid
phase is discovered by the perturbative renormalization group
equations. However, the model on an isotropic square lattice is
still unexplored. In particular, quantum Monte Carlo methods for
this model suffer the notorious sign problem except in the special
case where only the singlet bond exchange exists.

In this article, we present a systematic numerical study for the
$Sp(4)$ Heisenberg model at quarter filling in both 1D systems with
large sizes and 2D systems up to $4\times 4$ by means of exact
diagonalization techniques and the density matrix renormalization
group (DMRG)\cite{white1992,white1993}. In 1D, we numerically show
that the system exhibits two competing quantum phases: a
long-range-ordered gapped dimer phase when the exchange interaction
in the bond singlet channel ($J_0$) dominates over that in the
quintet channel ($J_2$), and a gapless spin liquid phase otherwise.
The $Sp(4)$ spin correlation functions are calculated, which shows
that in the dimer phase the correlations have the 2-site
periodicity, whereas in the gapless spin liquid phase they have the
4-sites periodicity. In 2D, our numerical simulations for small
sizes indicate three different dominant correlations depending on
the values of $J_0/J_2$. We infer three competing phases: the Neel
ordering, the plaquette ordering, and another possible phase of
columnar dimer ordering, in the thermodynamic limit.

The rest of this article is organized as follows. In Sec.
\ref{sect:model}, we introduce the Hamiltonian of spin-$\frac{3}{2}$
fermions which possesses the rigorous $Sp(4)$ symmetry, and a
magnetic exchange model in the Mott-insulating state at
quarter-filling. A self-contained introduction of the $Sp(4)/SO(5)$
algebra is given. Then we separate our main discussion into two
parts: Sec. \ref{sect:1D} for 1D and Sec. \ref{sect:2D} for 2D
systems. In Sec. \ref{sect:exact}, we study the low-energy spectra
of a finite size $Sp(4)$ chain with both open and periodic boundary
conditions. In Sec. \ref{sect:DMRG}, the DMRG calculation on the
spin correlation functions are presented to identify the gapped
$Sp(4)$ dimer phase and the gapless spin-liquid phase. In the second
half part, we first analyze the $2\times 2$ cluster in Sec.
\ref{sect:foursite} and perform exact diagonalization on larger
sizes to study the low-energy spectrum behavior in Sec.
\ref{sect:2Dspectra}. Then we display the calculations of the
magnetic structure form factor in Sec. \ref{sect:mag}, the dimer
correlation in Sec. \ref{sect:dimer} and the plaquette-type
correlation in Sec. \ref{sect:plaquette}. We discuss the possible
existence of the corresponding orderings. Conclusions are made in
the last section. At the end of this paper, we present a brief and
self-contained introduction to the representation theory of Lie
group in Appendix A to C.

\section{Model Hamiltonian and the hidden $Sp(4)$ symmetry}
\label{sect:model}

\subsection{The spin-$\frac{3}{2}$ Hubbard model}
We start with the generic one-band Hubbard model
loaded with spin-$\frac{3}{2}$ fermions. By neglecting long-range
Coulomb interactions, only onsite interactions are considered in the
Hubbard model. Due to Pauli's exclusion principle, the spin
wavefunctions of two onsite fermions have to be antisymmetric. The
total spin of two onsite spin-$\frac{3}{2}$ fermions can only be
either singlet $(S_T=0)$ or quintet $(S_T=2)$. We assign an
independent interaction parameter $U_0$ (singlet) and $U_2$
(quintet), respectively, to each channel. The Hamiltonian reads \bea
H&=&-t \sum_{\langle ij \rangle, \sigma} (\psi^{\dag}_{i \sigma}
\psi_{j \sigma}+h.c.)-\mu \sum_{i \sigma}\psi^{\dag}_{i \sigma}
\psi_{i \sigma} \nn
\\
&+&U_0 \sum_i P^{\dag}_{0}(i)P_{0}(i)
+U_2 \sum_{i,m=-2,..,2}P^{\dag}_{2m}(i)P_{2m}(i), \nn \\
\label{eq:hubbard} \eea where $\langle ij \rangle$ denotes the
nearest neighboring hopping; $\sigma$ represents four spin flavors
$F_z=\pm \frac{3}{2}$, $\pm \frac{1}{2}$; $P^{\dag}_{0}$ and
$P^{\dag}_{2,m}$ are the singlet and quintet pairing operators
defined through Clebsch-Gordon coefficients as \bea
P^{\dag}_{0}(i)&=&\sum_{\alpha\beta}\avg{00|\frac{3}{2}\frac{3}{2}\alpha\beta}
\psi^\dagger_\alpha(i) \psi^\dagger_\beta(i),\nn \\
P^{\dag}_{2m}(i)&=&\sum_{\alpha\beta}\avg{2m|\frac{3}{2}
\frac{3}{2}\alpha\beta} \psi^\dagger_\alpha(i) \psi^\dagger_\beta(i).
\eea

The actual symmetry of Eq. \ref{eq:hubbard} is much larger than the
$SU(2)$ symmetry: it has a hidden and exact $Sp(4)$, or,
isomorphically, $SO(5)$ symmetry. The $Sp(4)$ algebra can be
constructed as follows. For the 4-component fermions, there exist 16
bases for the $4\times 4$ Hermitian matrices $M_{\alpha\beta}
(\alpha,\beta= \pm\frac{3}{2}, \pm\frac{1}{2})$. They serve as
matrix kernels for the bi-linear operators, {\it i.e.},
$\psi_\alpha^\dagger M_{\alpha\beta}\psi_\beta$, in the
particle-hole channel. The density and 3-component spin $F_x, F_y,
F_z$ operators {\it do not} form a complete set. The other 12
operators are built up as high rank spin tensors, including
5-component spin-quadrupoles and 7-component spin-octupoles. The
matrix kernels of the spin-quadrupole operators are defined as \bea
\Gamma^1&=& \frac{1}{\sqrt 3} ( F_x F_y +F_y F_x ), \ \ \,
\Gamma^2= \frac{1}{\sqrt 3} ( F_z F_x +F_x F_z ),  \nn \\
\Gamma^3&=& \frac{1}{\sqrt 3} ( F_z F_y +F_y F_z ), \ \ \,
\Gamma^4= ( F_z^2-\frac{5}{4} ), \nn \\
\Gamma^5&=& \frac{1}{\sqrt 3} ( F_x^2 -F_y^2 ), \eea which
anti-commute with each other, and thus form a basis of the
Dirac-$\Gamma$ matrices. The matrix kernels of 3 spin and 7
spin-octupole operators together are generated from the commutation
relations among the 5 $\Gamma$-matrices as \bea \Gamma^{ab}=
-\frac{i}{2} [ \Gamma^a, \Gamma^b] \ \ \ (1\le a,b\le5). \eea
Consequently, these 16 bilinears can be classified as \bea
n&=&\psi^\dagger_\alpha\psi_\alpha, ~
n_a=\frac{1}{2}\psi^\dagger_\alpha \Gamma^a_{\alpha\beta}
\psi_\beta, ~ L_{ab} = -\frac{1}{2}\psi^\dagger_\alpha
\Gamma^{ab}_{\alpha\beta} \psi_\beta, ~~ \label{eq:generators} \eea
where $n$ is the density operator; $n_a$'s are 5-component
spin-quadrupole operators; $L_{ab}$'s  are 10-component spin and
spin-octupole operators \cite{WU2003,wu2006a}. Reversely the spin
$SU(2)$ generators $F_{x,y,z}$ can be written as
$F_+=\sqrt{3}(-L_{34}+i L_{24})-(L_{12}+iL_{25}) +i(L_{13}+iL_{35})$
and $F_z=L_{23}+2L_{15}$.

The 15 operators of $n_a$ and $L_{ab}$ together span the $SU(4)$
algebra. Among them, the 10 $L_{ab}$ operators are spin tensors with
odd ranks, and thus time-reversal (TR) odd, while the 5-component
$n_a$'s are TR even. The TR odd operators of $L_{ab}$ form a closed
sub-algebra of $Sp(4)$. The 4-component spin-$\frac{3}{2}$ fermions
form the fundamental spinor representation of the $Sp(4)$ group. In
contrast, the TR even operators of $n_a$ do not form a closed
algebra, but transform as a 5-vector under the $Sp(4)$ group. In
other words, $Sp(4)$ is isomorphic to $SO(5)$. But rigorously
speaking, the fermion spinor representations of $Sp(4)$ are not
representations of $SO(5)$. Their relation is the same as that
between $SU(2)$ and $SO(3)$. Below we will use the terms of $Sp(4)$
and $SO(5)$ interchangeably. The $SO(5)$ symmetry of Eq.
\ref{eq:hubbard} can be intuitively understood as follows. The
4-component fermions are equivalent to each other in the kinetic
energy term, which has an obvious $SU(4)$ symmetry. Interactions
break the $SU(4)$ symmetry down to $SO(5)$. The singlet and quintet
channels form the identity and $5$-dimensional vector
representations for the $SO(5)$ group, respectively, thus Eq.
\ref{eq:hubbard} is $SO(5)$ invariant without any fine-tuning.

\subsection{Magnetic exchanges at quarter-filling}
Mott-insulating states appear at commensurate fillings with strong
repulsive interactions. We focus on the magnetic exchange at quarter
filling, {\it i.e.}, one fermion per site. The Heisenberg type
exchange model has been constructed in Ref. [\onlinecite{chen2005}]
through the second-order perturbation theory. For each bond, the
exchange energies are $J_0=\frac{4t^2}{U_0}$ for the bond spin
singlet channel, $J_2=\frac{4t^2}{U_2}$ for the bond spin quintet
channel, and $J_1=J_3=0$ for the bond spin triplet and septet
channels, respectively. This exchange model can be written in terms
of bi-linear, bi-quadratic and bi-cubic Heisenberg exchange and the
Hamiltonian reads as
\bea
\label{eq:heisenberg}
H_{ex}=\sum_{\langle
i,j \rangle} a (\vec{F}_i\cdot \vec{F}_j) +b (\vec{F}_i \cdot
\vec{F}_j)^2+c (\vec{F}_i \cdot \vec{F}_j)^3,
\eea
where
$a=-\frac{1}{96}(31 J_0+23 J_2)$, $b=\frac{1}{72}(5 J_0 +17J_2)$ and
$c=\frac{1}{18}(J_0+J_2)$ and $F_{x,y,z}$ are usual $4\times 4$ spin
operators. Eq. \ref{eq:heisenberg} can be simplified into a more
elegant form with the explicitly $SO(5)$ symmetry \cite{wu2006a} as
\bea \label{eq:so5} H_{ex}&=&\sum_{\langle i,j \rangle} \Big\{
\sum_{1\le a < b \le 5} \frac{J_0+J_2}{4}
L_{ab}(i)L_{ab}(j)\nn \\
&+& \frac{3J_2-J_0}{4} \sum^5_{a=1} n_a(i)n_a(j) \Big \}. \eea In
the $SO(5)$ language, there are two diagonal operators commuting
with each other and read as \bea L_{15}&=&
\frac{1}{2}(n_{\frac{3}{2}}+n_{\frac{1}{2}}-n_{-\frac{1}{2}}
-n_{-\frac{3}{2}}), \nn \\
L_{23}&=& \frac{1}{2}(n_{\frac{3}{2}}-n_{\frac{1}{2}}
+n_{-\frac{1}{2}}-n_{-\frac{3}{2}}). \eea Corresponding to the spin
language, each singlet-site basis state can be labeled in terms of
these two quantum numbers as $|F_z \rangle = | L_{15},
L_{23}\rangle$: $|\pm \frac{3}{2} \rangle = | \pm \frac{1}{2}, \pm
\frac{1}{2}\rangle$ and $|\pm \frac{1}{2} \rangle = |\pm
\frac{1}{2}, \mp\frac{1}{2}\rangle$. For an arbitrary many-body
state, $L^{tot}_{15}=\sum_i L_{15}(i)$ and $L^{tot}_{23}=\sum_i
L_{23}(i)$ are good quantum numbers (similar to that
$F^{tot}_{z}=\sum_i F_{z}(i)$ is conserved in $SU(2)$ cases) and can
be applied to reduce dimensions of the Hilbert space in practical
numerical calculations.

There exist two different $SU(4)$ symmetries of Eq. \ref{eq:so5} in
two special cases. At $J_0=J_2=J$, {\it i.e.}, $U_0=U_2$, it reduces
to the $SU(4)$ Heisenberg model with each site in the fundamental
representation
\bea
H=\sum_{\langle i,j \rangle} \frac{J}{2} \Big\{
L_{ab}(i)L_{ab}(j)+n_a(i)n_a(j) \Big \}.
\eea
Below we denote this
symmetry as $SU(4)_A$. In this case, there is an additional good
quantum number $n_4$, \bea n_4= \frac{1}{2}
(n_{\frac{3}{2}}-n_{\frac{1}{2}}-n_{-\frac{1}{2}}
+n_{-\frac{3}{2}}). \eea This $SU(4)$ model is equivalent to the
Kugel-Khomskii type model \cite{kugel1982,sutherland1975} and is
used to study the physics with interplay between orbital and spin
degree of freedom.\cite{li1998,bossche2000,bossche2001}. On the
other hand, at $J_2=0$, {\it i.e.}, $U_2\rightarrow +\infty$, Eq.
\ref{eq:so5} has another $SU(4)$ symmetry in the bipartite lattice,
which is denoted $SU(4)_B$ below. In this case, we perform the
particle-hole transformation to one sublattice but leave the other
sublattice unchanged. The particle-hole transformation is defined as
$\psi_\alpha \rightarrow R_{\alpha\beta} \psi_\beta^\dagger$ where
$R$ is the charge conjugation matrix
\bea
R= \left (
\begin{array} {cc}
0 &i\sigma_2 \\
i\sigma_2 & 0 \\
\end{array}
\right ).
\label{eq:R4}
\eea
Under this operation, the fundamental representation transforms to
anti-fundamental representation whose $Sp(4)$ generators and vectors
become $L'_{ab}=L_{ab}$ and $n'_{a}=-n_{a}$.
Thus Eq. \ref{eq:so5} can be recast to
\bea
H=\sum_{\langle i,j\rangle}\frac{J}{2}\big(L'_{ab}(i)L_{ab}(j)+n'_a(i)n_a(j)
\big),
\label{eq:SU4B}
\eea
which is $SU(4)$ invariant again.

These two $SU(4)$ symmetries have very different physical
properties. In the case of $SU(4)_A$, two sites are not enough to
form an $SU(4)$ singlet. It at least needs four sites to form an
$SU(4)$ singlet as
$\epsilon_{\alpha\beta\gamma\delta}\psi^\dagger_\alpha(1)\psi^\dagger_\beta(2)
\psi^\dagger_\gamma(3)\psi^\dagger_\delta(4)$, where
$\epsilon_{\alpha\beta\gamma\delta}$ is the rank-4 fully
antisymmetric tensor. Thus quantum magnetism of Eq. \ref{eq:so5} at
$J_0=J_2$ is characterized by four-site correlations. The ground
state of such a system on a 2D square lattice was conjectured to be
a plaquette $SU(4)$ singlet state without magnetic long-ranged
ordering.\cite{bossche2000,Mishra2002} On the other hand, for the
$SU(4)_B$ case, two sites can form an $SU(4)$ singlet as
$R_{\alpha\beta}\psi^\dagger_\alpha(1) \psi^\dagger_\beta(2)$. In
the 2D square lattice, a long-ranged Neel order is identified by
quantum Monte Carlo simulations\cite{harada2003} and large $N$
limit.\cite{Read1990} The square of the staggered magnetization is
numerically given as $m_s=0.091$, which is much smaller than that of
the $SU(2)$ Neel order state.

\section{Quantum magnetism in the 1D chains}
\label{sect:1D}

\begin{figure}[htb]
\centering\epsfig{file=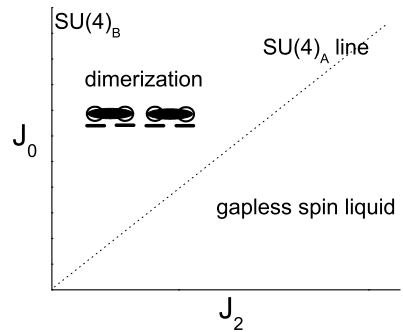,clip=1,width=0.6\linewidth,angle=0}
\caption{Phase diagram of the 1D chain in terms of the singlet and
quintet channel interaction $J_0$ and $J_2$. In this context,
$\theta$ is the angle defined by $\theta=\tan^{-1}(J_0/J_2)$. The
$SU(4)_A$ type ($\theta=45^{\circ}$) denoted by the dot line belongs
to the gapless spin liquid state whereas $SU(4)_B$ along $J_2=0$.
The phase boundary separating the dimerization phase and the gapless
liquid state is the $SU(4)_A$ line.} \label{fig:phase}
\end{figure}

We start our discussion on the 1D chain. The phase diagram of the 1D
spin-$\frac{3}{2}$ Hubbard model has been studied by one of the
author using the method of bosonization \cite{WU2005a,wu2006a}. At
the commensurate quarter-filling (one particle per site) with purely
repulsive interactions $(U_0>0, U_2>0)$, the $4k_f$-Umklapp term
opens a charge gap as $K_c<\frac{1}{2}$. In this case,  the physics
is captured by the exchange model of Eq. \ref{eq:so5}. It has been
found that in the regime of $J_0/J_2>1$ dimerization of spin
Perierls order is present, whereas it is a gapless spin liquid phase
at $J_0/J_2 \le 1$ (see Fig. \ref{fig:phase}) \cite{WU2005a}. In the
following, we use exact diagonalization methods and DMRG not only
identify these two competing phases but also demonstrate the ground
state profiles and 4-site periodicities in spin-spin correlations.

\subsection{Exact diagonalization on low energy spectra}
\label{sect:exact}

In this subsection, we apply the exact diagonalization technique to
study the 1D $Sp(4)$ spin-$\frac{3}{2}$ chains with nearest neighbor
exchange interactions described by Eq. \ref{eq:so5}. We only
consider the case of the site number $N=4m$. For convenience, we set
$J_0=\sqrt{2}\sin \theta$ and $J_2=\sqrt{2} \cos \theta$. Regardless
of $\theta$ and sizes $N$, the ground states (GS) only exist in the
$(L^{tot}_{15},L^{tot}_{23})=(0,0)$ sector and are unique with
$C=0$, where $C$ denotes the $Sp(4)$ Casimir of the entire system
and is expressed in terms of the $Sp(4)$ generators as \bea C=
\sum_{1\le a<b \le 5}\Big\{\sum_{i} L_{ab} (i) \Big \}^2. \eea In
addition to $L^{tot}_{15}$ and $L^{tot}_{23}$, the Casimir $C$ is
also a conserved quantity in the $Sp(4)$ system, analogous to the
total spin in $SU(2)$ systems. Each energy eigenstate can be labeled
by $C$ and further identified the dimension of the representation
(degeneracy). As shown in the table II in the Appendix, while $C=0$,
the state is an $Sp(4)$ singlet and unique whereas while $C > 0$ the
state is multiplet and has degeneracy which is equal to the
dimension of the associated representation.

\begin{figure}[htb]
\centering\epsfig{file=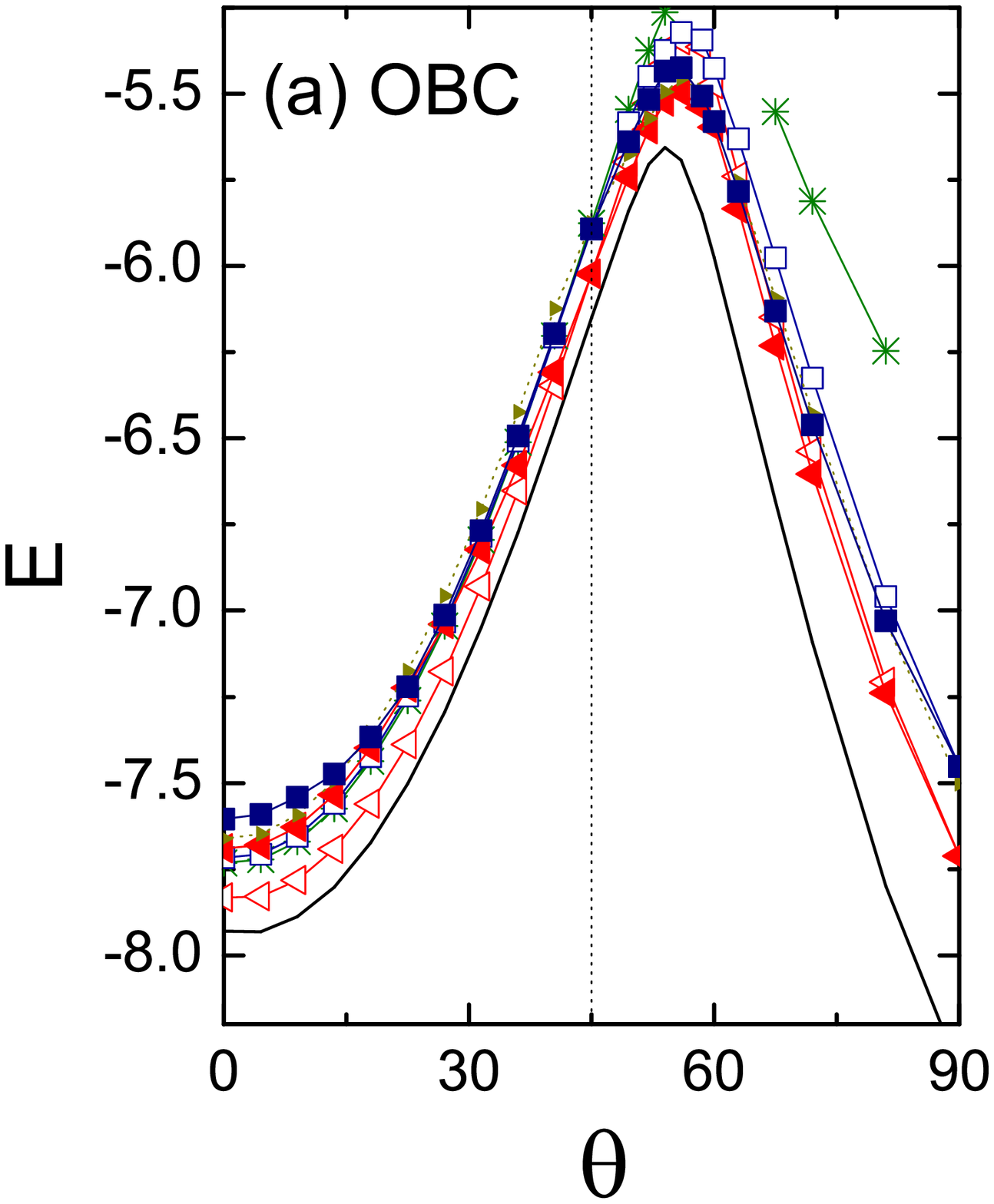,clip=1,width=0.43\linewidth,angle=0}
\centering\epsfig{file=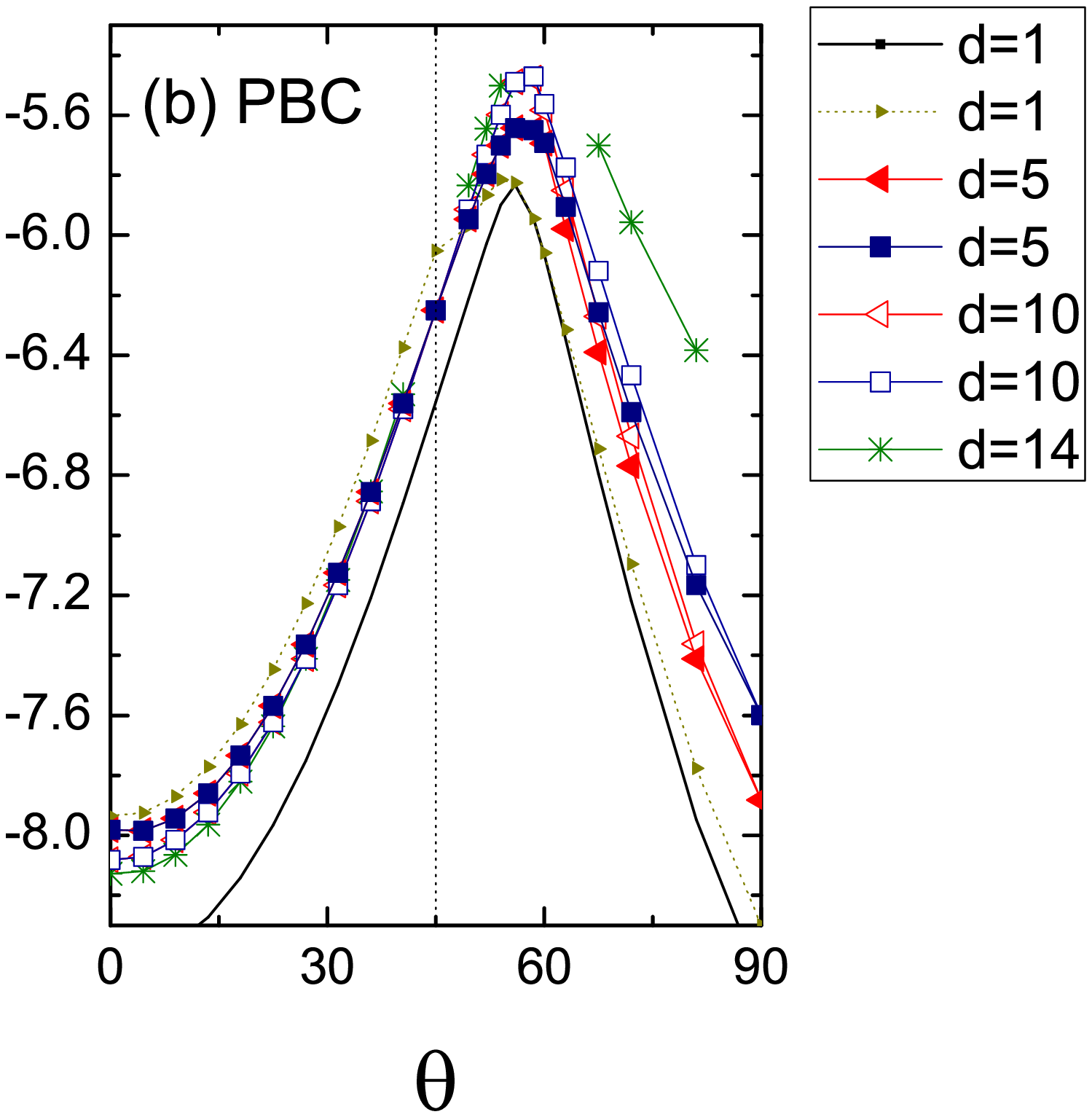,clip=1,width=0.52\linewidth,angle=0}
\caption{The exact diagonalization on 1D chain with 12 sites for (a)
open and (b) periodic boundary conditions. The dispersion of the
ground state and low excited states, and the dimensions d of their
corresponding representations of the $Sp(4)$ group are shown. }
\label{fig:exact_8sites}
\end{figure}
In Fig. \ref{fig:exact_8sites} (a) and (b), the ground state and low
excited states with 12 sites are presented using open and periodic
boundary conditions, respectively. The GS as varying $\theta$ angles
is always an $Sp(4)$ singlet, which becomes an $SU(4)$ singlet at
$\theta=45^\circ$ ($SU(4)_A$) and $\theta=90^\circ$ ($SU(4)_B$) for
both boundary conditions. For the low energy excited states, we
first look at the regime of $45^\circ<\theta<90^\circ$, {\it i.e.},
$J_0>J_2$. With open boundary conditions (OBC), the lowest excited
states (LES) are the $Sp(4)$ 5-vector states  with the quadratic
Casimir $C=4$. The next lowest excited states (NLES) are 10-fold
degenerate and belong to the 10-dimensional ($10d$) $Sp(4)$ adjoint
representation with $C=6$. The LES and NLES merge at both of the
$SU(4)_A~(\theta=45^\circ)$ and $SU(4)_B ~(\theta=90^\circ)$
points, and become 15-fold degenerate. This is the $SU(4)$ adjoint
representation with $C=8$. With periodic boundary conditions (PBC),
the 5-vector and the 10-fold states behave similarly as before.
However, a marked difference is that a new $Sp(4)$ singlet state
appears as the LES at $50^\circ<\theta<90^\circ$, which becomes
higher than the 5-vector states only very close to $45^\circ$. In
particular, it is nearly degenerate with the ground state (which is
the lowest $Sp(4)$ singlet) at $\theta=50^{\circ} \sim 60^{\circ}$.
In the regime of $0^\circ<\theta<45^\circ$, {i.e.}, $J_2>J_0$ the
excited states are many $Sp(4)$ multiplets with energies close to
each other. With OBC, the LESs form the 10d $Sp(4)$ adjoint
representation. For the PBC case, the 14-dimensional symmetric
tensor representation of $Sp(4)$ competes with the 10d adjoint one.

The appearance of two nearly degenerate $Sp(4)$ singlets at
$50^\circ<\theta<90^\circ$ with PBC and their disappearance with OBC
can be understood by the dimerization instability. The dimerization
and the spin gapped ground state was shown in the bosonization
analysis at $45^\circ<\theta<90^\circ$ \cite{WU2005a}. In the
thermodynamic limit, the ground state has double degeneracy
corresponding to two different dimer configurations, both
spontaneously breaking translational symmetry. The OBC favors only
one of the dimer configurations, but disfavors the other due to one
bond breaking. In the finite system with PBC, the two dimer
configurations tunnel between each other, which gives rise to two
nearly degenerate $Sp(4)$ singlet states. We further calculate the
gap between them, denoted by $\Delta_{ss}$, at $\theta
> 45^{\circ}$ by using exact diagonalization under PBC up to 16
sites.

\begin{figure}
\centering\epsfig{file=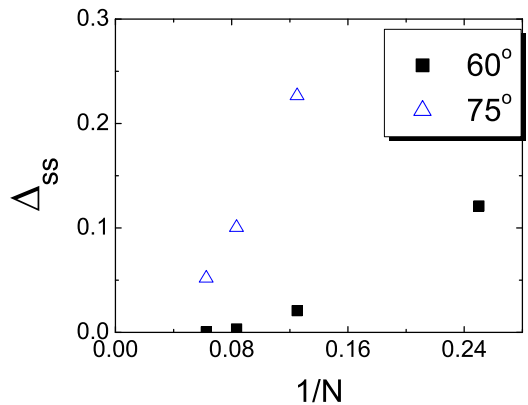,clip=1,width=0.8\linewidth,angle=0}
\caption{ Exact diagonalization results on the $Sp(4)$
singlet-singlet gap with $J_0>J_2$ and periodic boundary conditions
( $\theta=60^{\circ}$ and $75^{\circ}$ with $N=8,12$ and $16$).
Finite size scaling shows the vanishing of the singlet-singlet gap
$\Delta_{ss}$.} \label{fig:gap_period}
\end{figure}

As presented in Fig. \ref{fig:gap_period}, $\Delta_{ss}$ disappears
in the finite size scaling due to the twofold degeneracy. On the
other hand, the existence of the spin gap in this parameter regime
is presented in Fig. \ref{fig:gap_open} by DMRG simulation in Sec.
\ref{sect:DMRG} below. The original Lieb-Schultz-Mattis theorem
\cite{lieb1961} was proved that for the $SU(2)$ case, the GS of
half-integer spin chains with translational and rotational
symmetries is gapless, or gapped with breaking translational
symmetry. It is interesting to observe that our results of the
$Sp(4)$ spin chain also agree with this theorem. The nature of the
GS in the parameter regime $0^\circ<\theta<45^\circ$ will be
discussed in Sec. \ref{sect:DMRG}.

\subsection{DMRG simulations on $Sp(4)$ spin chain} \label{sect:DMRG}
\label{sect:DMRG1D}

\begin{figure}[htb]
\centering\epsfig{file=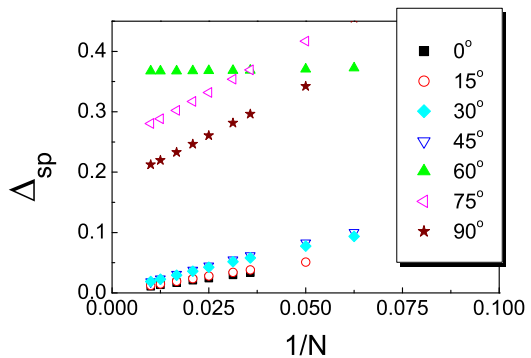,clip=1,width=0.8\linewidth,angle=0}
\caption{The finite size scaling of the spin gap $\Delta_{sp}$ of
the $Sp(4)$ spin chain vs $1/N$ at various values of $\theta$.
$\theta$ is defined as $\theta=\tan^{-1}(J_0/J_2)$ and $N$ is the
system size.} \label{fig:gap_open}
\end{figure}

In this subsection, we present the DMRG calculations on the ground
state properties of the $Sp(4)$ chain up to 80 sites with OBC. We
first present the spin gap $\Delta_{sp}$ in Fig. \ref{fig:gap_open},
which is defined as the energy difference between the ground state
and the lowest $Sp(4)$ multiplet.

For chains with even number of sites, the GS is obtained with
quantum number $L^{tot}_{15}=L^{tot}_{23}=0$, and any $Sp(4)$
multiplet contains the states with quantum numbers
$(L^{tot}_{15}=\pm 1, L^{tot}_{23}=0)$ and $(L^{tot}_{15}=0,
L^{tot}_{23}=\pm 1)$. States with the same values of $(L^{tot}_{15},
L^{tot}_{23})$ may belong to different $Sp(4)$ representations,
which can be distinguished by their $Sp(4)$ Casimir. Practically, we
only need to calculate these sectors with low integer values of
$(L^{tot}_{15}, L^{tot}_{23})$ to determine the spin gaps. For the
cases of $\theta
> 45^{\circ}$, i.e., ($J_2/J_0 < 1$), $\Delta_{sp}$s saturate to
nonzero values as $1/N \to 0$, indicating the opening of spin gaps.
On the other hand, $\Delta_{sp}$'s vanish at $\theta \le 45^\circ$,
which shows that the ground state is gapless. These results agree
with the bosonization analysis \cite{WU2005a}, which shows that the
phase boundary is at $\theta=45^\circ$ with the $SU(4)_A$ symmetry,
which is also gapless. This gapless $SU(4)_A$ line was also studied
before in Ref. [\onlinecite{yamashita1998,azaria1999}].

\begin{figure}[!htb]
\centering\epsfig{file=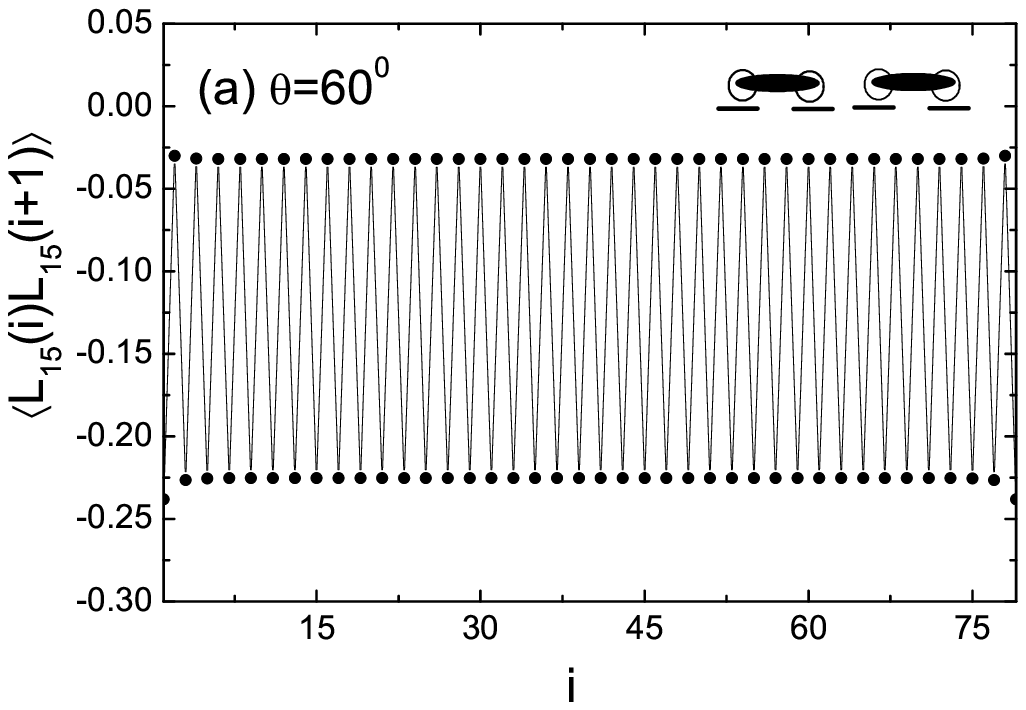,clip=1,width=0.7\linewidth,angle=0}
\centering\epsfig{file=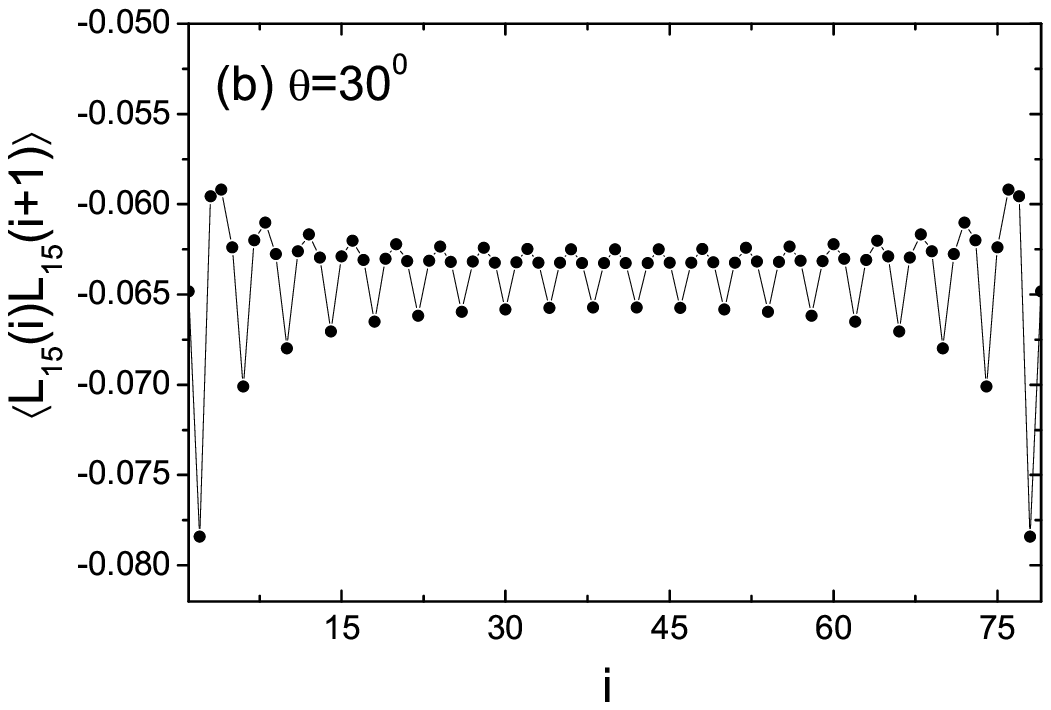,clip=1,width=0.73\linewidth,angle=0}
\caption{The NN correlation of $\langle L_{15}(i)L_{15}(i+1)
\rangle$ with open boundary conditions for (a) $\theta=60^{\circ}$
and (b) $\theta=30^{\circ}$, respectively. The dimer ordering is
long-ranged in (a). Note the 2-site periodicity in (a) and the
4-site periodicity in (b).} \label{fig:nncorr}
\end{figure}

To further explore the ground state profile, we calculate the
nearest neighbor (NN) correlation functions of the $Sp(4)$
generators for a chain of 80 sites. This correlation function is
similar to the bonding strength and defined as $\langle X(i)
X(i+1)\rangle$, where  $X$ are $Sp(4)$ generators. We present the
result of $\langle L_{15}(i) L_{15}(i+1)\rangle$ in Fig.
\ref{fig:nncorr}, and the correlation functions of other generators
should be the same due to the $Sp(4)$ symmetry. The open boundary
induces characteristic oscillations. At $\theta=60^\circ$, {\it
i.e.}, $J_0/J_2=\sqrt{3}$, $\langle L_{15}(i) L_{15}(i+1)\rangle$
exhibits the dominant dimer pattern, which does not show noticeable
decay from the edge to the middle of the chain. This means that the
dimerization is long-range-ordered in agreement with the
bosonization analysis \cite{wu2006a}. In contrast, at
$\theta^\circ=30^\circ$, {\it i.e.}, $J_2/J_0=\sqrt 3$, $\langle
L_{15}(i) L_{15}(i+1)\rangle$ exhibits a characteristic power-law
decay with 4-site periodicity oscillations. The 4-site periodicity
is also observed at other $\theta$'s for $\theta\le 45^\circ$, same
as ones presented in the bosonization analysis.

\begin{figure}[!htb]
\centering\epsfig{file=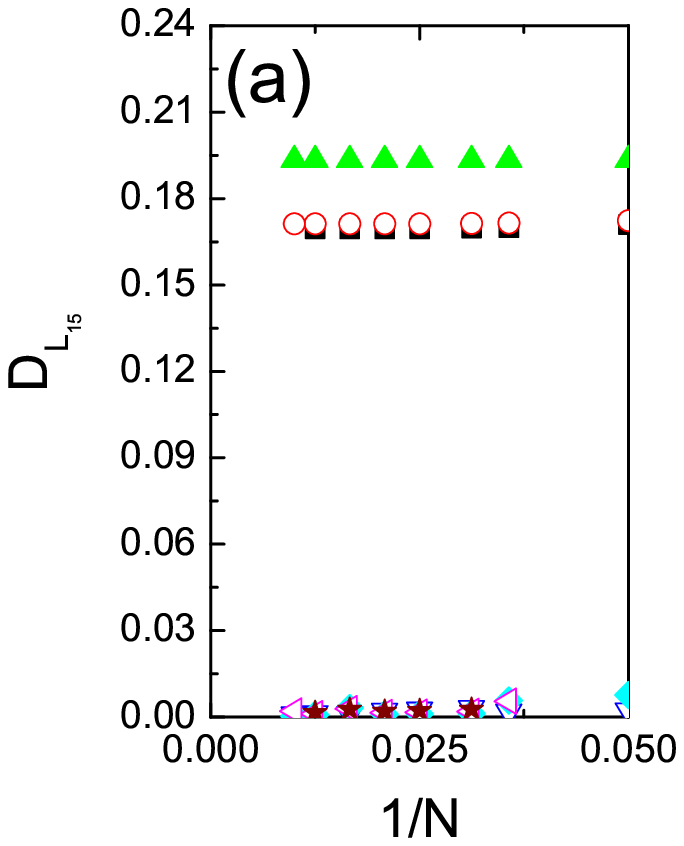,clip=1,width=0.4\linewidth,angle=0}
\centering\epsfig{file=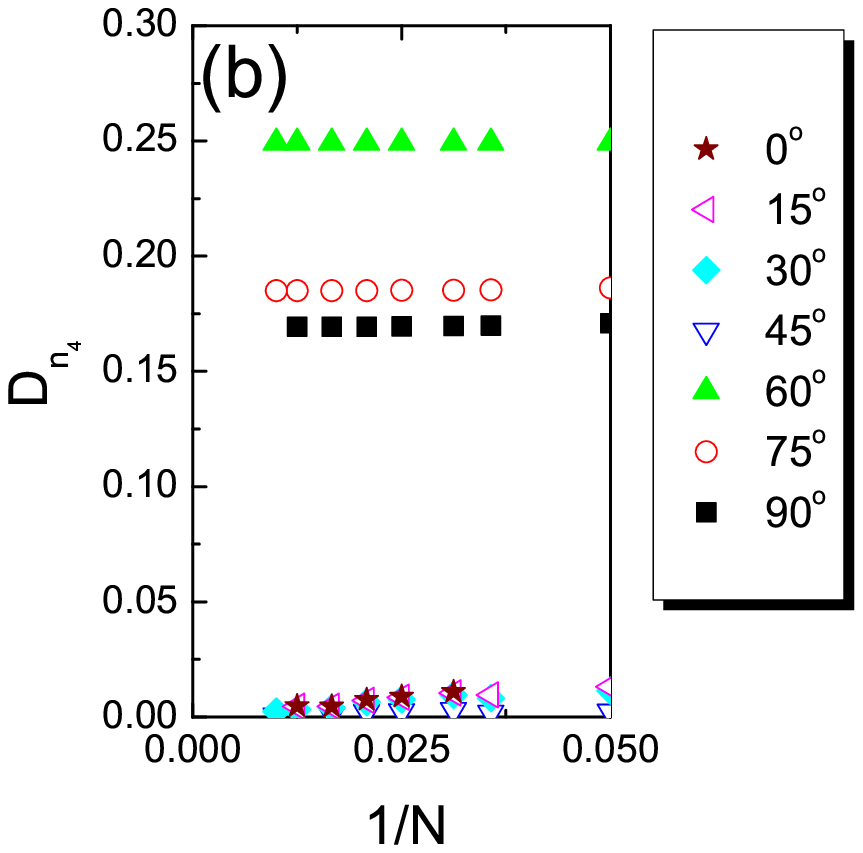,clip=1,width=0.5\linewidth,angle=0}
\caption{The finite size scaling for the dimer order parameters (a)
$D_{L_{15}}$ and (b) $D_{n_4}$ vs $1/N$ at various $\theta$'s.}
\label{fig:dimerp}
\end{figure}

We follow the definition for the dimer order parameter in Ref.
[\onlinecite{fath2001,hung2006}] as \bea D_{X}=|\langle
X(\frac{N}{2}-1) X(\frac{N}{2})\rangle-\langle
X(\frac{N}{2})X(\frac{N}{2}+1)\rangle|. \nn \\
\label{eq:dimer} \eea As shown previously, $X$s are $Sp(4)$
generators and vectors in the $Sp(4)$ spin chain. Without loss of
generality, we choose two non-equivalent operators as $X=L_{15}$ and
$n_4$ for $Sp(4)$ generators and vectors, respectively. The open
boundary conditions provide an external field to pin down the dimer
orders. The finite size scaling of the dimer orders of the two
middle bonds is presented in Fig. \ref{fig:dimerp} (a) and (b) at
various values of $\theta$, respectively. It is evident that in the
regime of $\theta
> 45^{\circ}$ both of $D_{L_{15}}$ and $D_{n_4}$ remain finite as
$1/N \to 0$ whereas for $\theta \le 45^{\circ}$ the dimer order
parameters vanish. We conclude that the ground state is the dimer
phase for  $J_0/J_2 > 1$.

\begin{figure}
\centering\epsfig{file=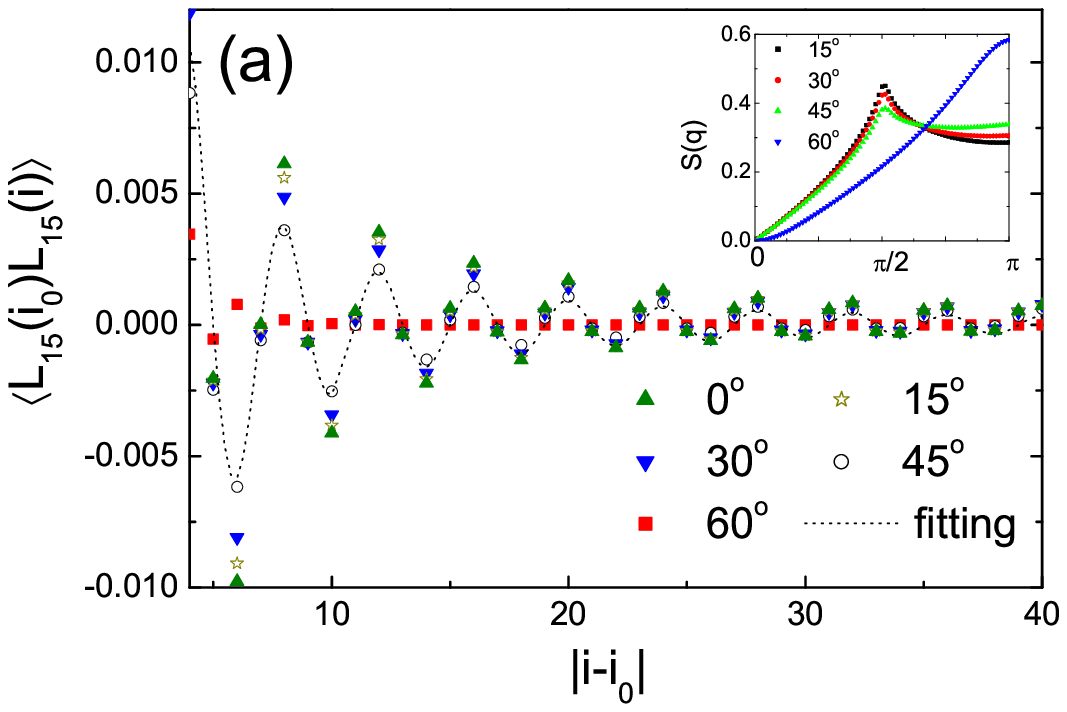,clip=1,width=0.9\linewidth,angle=0}
\centering\epsfig{file=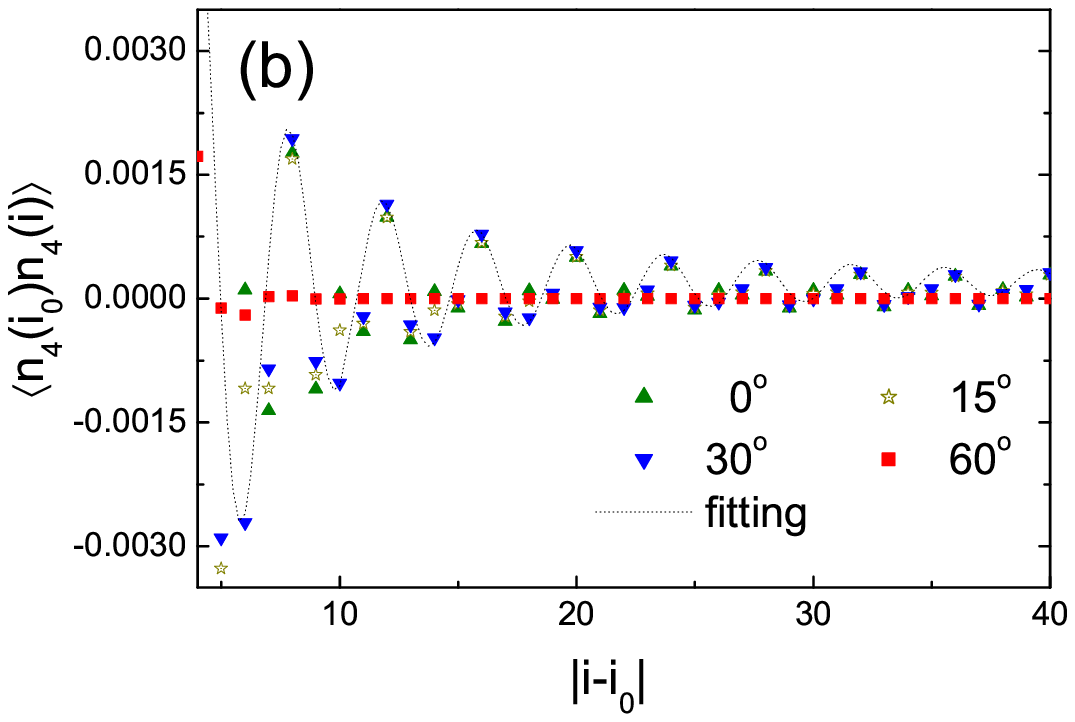,clip=1,width=0.9\linewidth,angle=0}
\caption{(a) The two point correlations $\langle
L_{15}(i_0)L_{15}(i) \rangle$ at $\theta=0^{\circ}$,$15^{\circ}$,
$30^{\circ}$, $45^{\circ}$ and $60^{\circ}$. The dot line is plotted
by the fitting result using $cos(x \pi /2)/x^{1.52}$. The reference
point $i_0$($=40$) is the most middle site of the chain ($N=80$).
The inset indicates that all $S(q)$ for $\theta\le 45^{\circ}$ have
peaks located at $q=41\pi/81 \sim \pi/2$ whereas $\pi$ for
$\theta=60^{\circ}$. (b) $\langle n_{4}(i_0)n_{4}(i) \rangle$ at
$\theta=0^{\circ}$,$15^{\circ}$, $30^{\circ}$ and $60^{\circ}$ and
the fitting uses $\kappa=1.55$.}
\label{fig:corr}
\end{figure}

Next we present the two point correlation functions of $\langle X(i)
X(j) \rangle$, where $X$ is $L_{15}$ and $n_4$, in Fig.
\ref{fig:corr} (a) and (b), respectively. At $\theta>45^\circ$, say,
$\theta=60^\circ$, both correlation functions show exponential decay
due to the dimerization. In the spin liquid regime of $\theta\le
45^\circ$, {\it i.e.} $J_2\ge J_0$, however, all the correlation
functions exhibit the power-law behavior and the same $2k_f$
oscillations with the 4-site period. Their asymptotic behavior can
be written as \bea \langle  X(i_0) X(i) \rangle \propto \frac{\cos
\frac{\pi}{2} |i-i_0|} {|i-i_0|^\kappa}. \eea  Along the $SU(4)_A$
line ($\theta=45^\circ$), the correlations of $L_{15}$ and $n_4$ are
degenerate. The power can be fitted as $\kappa\approx 1.52$, which
is in good agreement with the value of $1.5$ from bosonization
analysis and numerical studies
\cite{affleck1986,WU2005a,yamashita1998,azaria1999}. As $\theta$ is
away from $45^\circ$, the $SU(4)$ symmetry is broken. For the
correlations of $L_{15}$, the values of $\kappa$ decrease as
decreasing $\theta$, which can be fitted as $\kappa=1.41, 1.34,
1.30$ for $\theta=30^\circ, 15 ^\circ, 0^\circ$, respectively. On
the other hand, for the correlations of $n_4$, the values of
$\kappa$ can be fitted as $\kappa=1.55, 1.65, 1.60$ for
$\theta=30^\circ, 15 ^\circ, 0^\circ$, respectively. We also perform
the Fourier transforms of the correlations of $\langle
L_{15}(i_0)L_{15}(i) \rangle$, $S(q)$, and present the results in
the inset of Fig. \ref{fig:corr} (a). $S(q)$ is defined as \bea
S(q)=\sum_{i,j}e^{i q(r_i-r_j)}\langle L_{15}(r_i)L_{15}(r_j)
\rangle \eea and $q=m\pi/(N+1)$, where $m=1,2 \cdots, N$ are
integers for OBC. Clearly, in the regime of $\theta \le 45^{\circ}$
all the peaks are located at $q=41\pi/81 \sim \pi/2$, indicating a
$2k_f$ charge density wave. On the other hand, $S(q)$ at
$\theta=60^{\circ}$ appears a peak at $\pi$, which denotes a $4k_f$
charge density wave and is characteristic of the dimerization phase.


\section{The $Sp(4)$ magnetism in 2D square lattice with small sizes}
\label{sect:2D}

\begin{figure} [!hbt]
\centering\epsfig{file=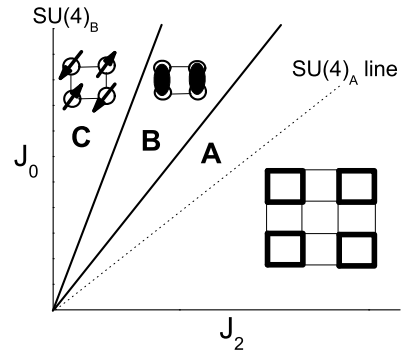,clip=1,width=0.6\linewidth,angle=0}
\caption{Speculated phase diagram of the 2D $Sp(4)$ spin-$3/2$
systems at quarter filling from Ref. [\onlinecite{wu2006a}].
$\theta=\tan^{-1}(J_0/J_2)$. The
$SU(4)_A$ type drawn by the dot line is at $J_0=J_2$
($\theta=45^{\circ}$) whereas $SU(4)_B$ at $J_2=0$
($\theta=90^{\circ}$). Bold letters {\bf A}, {\bf B}, and {\bf C}
represent the plaquette, columnar dimerized and Neel order states,
respectively.}
\label{fig:2Dphasediagram}
\end{figure}

The quantum magnetism of Eq. \ref{eq:so5} in 2D is a very
challenging problem. Up to now, a systematic study  is still void.
In the special case of the $SU(4)_B$ line, i.e. $J_2=0$, in the
square lattice, quantum Monte-Carlo simulations are free of the sign
problem, which shows the long-range-Neel ordering but with very
small Neel moments $n_4 =(-)^i L_{15} = (-)^i L_{23}\approx 0.05$
\cite{harada2003}. This result agrees with the previous large-$N$
analysis \cite{zhang2001}. The Goldstone manifold is
$CP(3)=U(4)/[U(1) \otimes U(3)]$ with 6 branches of spin-waves. On
the other hand, on the $SU(4)_A$ line with $J_0=J_2$, an exact
diagonalization study on the $4\times 4$ sites shows the evidence of
the four-site $SU(4)$ singlet plaquette ordering \cite{bossche2000}.
Large size simulations are too difficult to confirm this result. On
the other hand, a variational wavefunction method based on the
Majorana representation of  spin operators suggests a spin-liquid
state at the $SU(4)_A$ line \cite{wang2009}. Recently, Chen {\it et
al.} \cite{chen2005} constructed an $SU(4)$ Majumdar-Ghosh model in
a two-leg spin-3/2 ladder whose ground state is solvable exhibiting
this plaquette state. An $SU(4)$ resonant plaquette model in 3D have
also been constructed \cite{xu2008,pankov2007}.

Based on these available knowledge, a speculated phase diagram was
provided in Ref. [\onlinecite{wu2006a}] as shown in Fig.
\ref{fig:2Dphasediagram}.
The Neel order state {\bf C} is expected to extend to a region with
finite $J_2$ instead of only along the $J_2=0$ line.
Furthermore, the plaquette order phase {\bf A} exists not only along the
$SU(4)_A$ line but also covers a finite range including
$\theta=45^{\circ}$.
Between {\bf A} and {\bf C}, there exists an intermediate phase {\bf B}
which renders ordered dimerizations which are two-sites spin singlets.
However, these features have not been tested due to the lack of
controllable analytic and numeric methods for 2D strongly
correlation systems.
For example, quantum Monte Carlo methods suffer notorious
sign problems at $J_2\neq 0$.

In this section, we begin with the cluster of
$2\times 2$ whose ground states can be solved analytically.
Then we perform exact diagonalization (ED) methods for the case
of $4\times 4$ sites and analyze the associated GS profiles for
different values of $\theta$.
Even though the size that we are studying is still small to draw any
conclusion for the thermodynamic limit, it provides valuable information
on the ground state properties.


\subsection{The $2 \times 2$ cluster}
\label{sect:foursite}

\begin{figure}[!htb]
\centering\epsfig{file=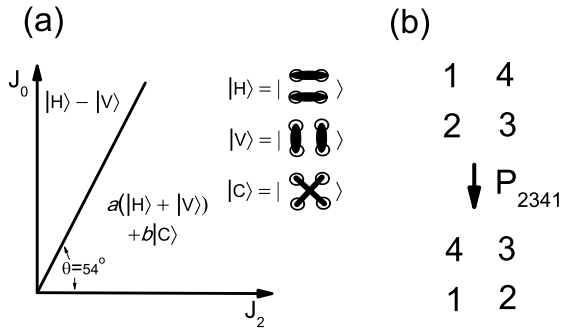,clip=1,width=0.85\linewidth,angle=0}
\caption{(a) The GS wavefunctions of the $2 \times 2$ cluster at
various $\theta$. $a$ and $b$ are coefficients depending on $\theta$
and the thick bonds denote the two-site $Sp(4)$ spin singlet states.
(b) The position indices before and after the permutation
$P_{2341}$.} \label{fig:2x2cluster}
\end{figure}

We begin  with the $2 \times 2$ cluster, whose ground states can be solved
analytically for all the values of $\theta$.
Such a system contains three $Sp(4)$ singlets, and the
ground states can be expanded in this singlet subspace.
These $Sp(4)$ singlets can be conveniently represented in terms
of the dimer states with the horizontal, vertical, and
cross diagonal configurations
depicted in Fig. \ref{fig:2x2cluster} (a) as
\bea
|H\rangle&=&\frac{1}{4}R_{\alpha\beta} \psi^\dagger_\alpha(4)
\psi^\dagger_\beta(1)
R_{\gamma\delta}\psi^\dagger_\gamma(2) \psi^\dagger_\delta(3) |\Omega\rangle, \nn \\
|V\rangle&=&\frac{1}{4}R_{\alpha\beta} \psi^\dagger_\alpha(1)
\psi^\dagger_\beta(2)
R_{\gamma\delta}\psi^\dagger_\gamma(3) \psi^\dagger_\delta(4) |\Omega\rangle, \nn \\
|C\rangle&=&\frac{1}{4}R_{\alpha\beta} \psi^\dagger_\alpha(1)
\psi^\dagger_\beta(3) R_{\gamma\delta}\psi^\dagger_\gamma(2)
\psi^\dagger_\delta(4) |\Omega\rangle,
\eea
where $R$ is the charge
conjugation matrix define in Eq. \ref{eq:R4}.
These states are linearly independent but are not orthogonal to each
other, satisfying $\avg{H|V}=\avg{V|C}=\avg{C|H}=-\frac{1}{4}$.
Under the permutation of the four sites $P_{(2341)}$, or a rotation
at $90^{\circ}$ as shown in Fig. \ref{fig:2x2cluster} (b), they
transform as
\bea
P_{2341}|H\rangle =|V\rangle, ~ P_{2341}|V\rangle
=|H\rangle, ~ P_{2341}|C\rangle =|C\rangle.
\eea

At $\theta=45^{\circ}$, i.e., the $SU(4)_A$ case, the ground state (GS) is
exactly an $SU(4)$ singlet over the sites $1$ to $4$:
\cite{li1998,chen2005}
\bea \label{eq:su4singlet}
|\Psi^s_{SU(4)}\rangle=\frac{1}{\sqrt{4!}} \sum_{\mu \nu \tau \xi}
\varepsilon_{\mu \nu \tau \xi}
\psi^{\dag}_{\mu,1}\psi^{\dag}_{\nu,2}\psi^{\dag}_{\tau,3}\psi^{\dag}_{\xi,4}
|\Omega \rangle, \eea
where the indices
$\mu$,$\nu$,$\tau$,$\xi$ run over $\pm \frac{3}{2},\pm \frac{1}{2}$;
$|\Omega \rangle$ represents the vacuum state; $\varepsilon_{\mu \nu
\tau \xi}$ is a rank-four fully antisymmetric tensor.
It can also be represented as the linear combination of the dimer states as
\bea
|\Psi^s_{SU(4)}\rangle=\sqrt{\frac{2}{3}} \big(|H\rangle +|V\rangle+
|C\rangle \big),
\eea
which is even under the rotation operation $P_{2341}$.
We find that in the entire range of
$0\le\theta<54^\circ$, the GS wavefunctions remain even under such
a rotation $P_{2341}$, whose wavefunctions can be represented as
\bea
|\Psi\rangle= a\big(|H\rangle +|V\rangle\big)+ b|C\rangle,
\label{eq:GSWF}
\eea
where $a$ and $b$ are coefficients depending on
the values of $\theta$. In fact, the overlaps between GS
wavefunctions Eq. \ref{eq:GSWF} and the $SU(4)$ singlet state $|
\Psi_{SU(4)}\rangle$ are larger than $0.98$ at $\theta<54^\circ$.
At $\theta>54^{\circ}$, a level crossing occurs and the
GS wavefunction changes to
\bea
|\Psi^s_{SU(4)}\rangle= \sqrt{\frac{2}{3}} \Big( |H\rangle
-|V\rangle \Big),
\eea
which is independent of $\theta$ and odd
odd under the rotation $P_{2341}$.

Combining the above observations, we identify that there are two
competing states in the system.
The boundary is located at $\theta=54^{\circ}$.
Next we turn to analyze large size systems.

\subsection{The Low energy spectra for the $4\times 4$ cluster}
\label{sect:2Dspectra}

In this subsection we study a larger system size of $N=4\times 4$.
Both $L^{tot}_{15}=\sum_i L_{15}(i)$ and $L^{tot}_{23}=\sum_i L_{23}(i)$
are good quantum numbers, which can used to reduce the Hilbert space.
The dimension of the Hilbert space
in the $(L^{tot}_{15},L^{tot}_{23})=(0,0)$ sector goes up to $165$ million.
On the other hand, the lowest multiplet states are located
in the sector of $(L^{tot}_{15},L^{tot}_{23})=(0,\pm 1)$ or
$(\pm 1,0)$ and the corresponding dimension is about $147$ million.
The dimensions of the subspace are too large to perform
diagonalization.
Nevertheless, by using translational symmetry, the dimension of the
Hilbert space reduces to $10$ million such that ED calculations
become doable.
The ground states are always in the sector of total momentum
$\vec K=(0,0)$, and are $Sp(4)$ singlets.
In the following, except for the specific mention in Sec.
\ref{sect:plaquette}, the systems are considered under periodic
boundary conditions.

\begin{figure}[!htb]
\centering\epsfig{file=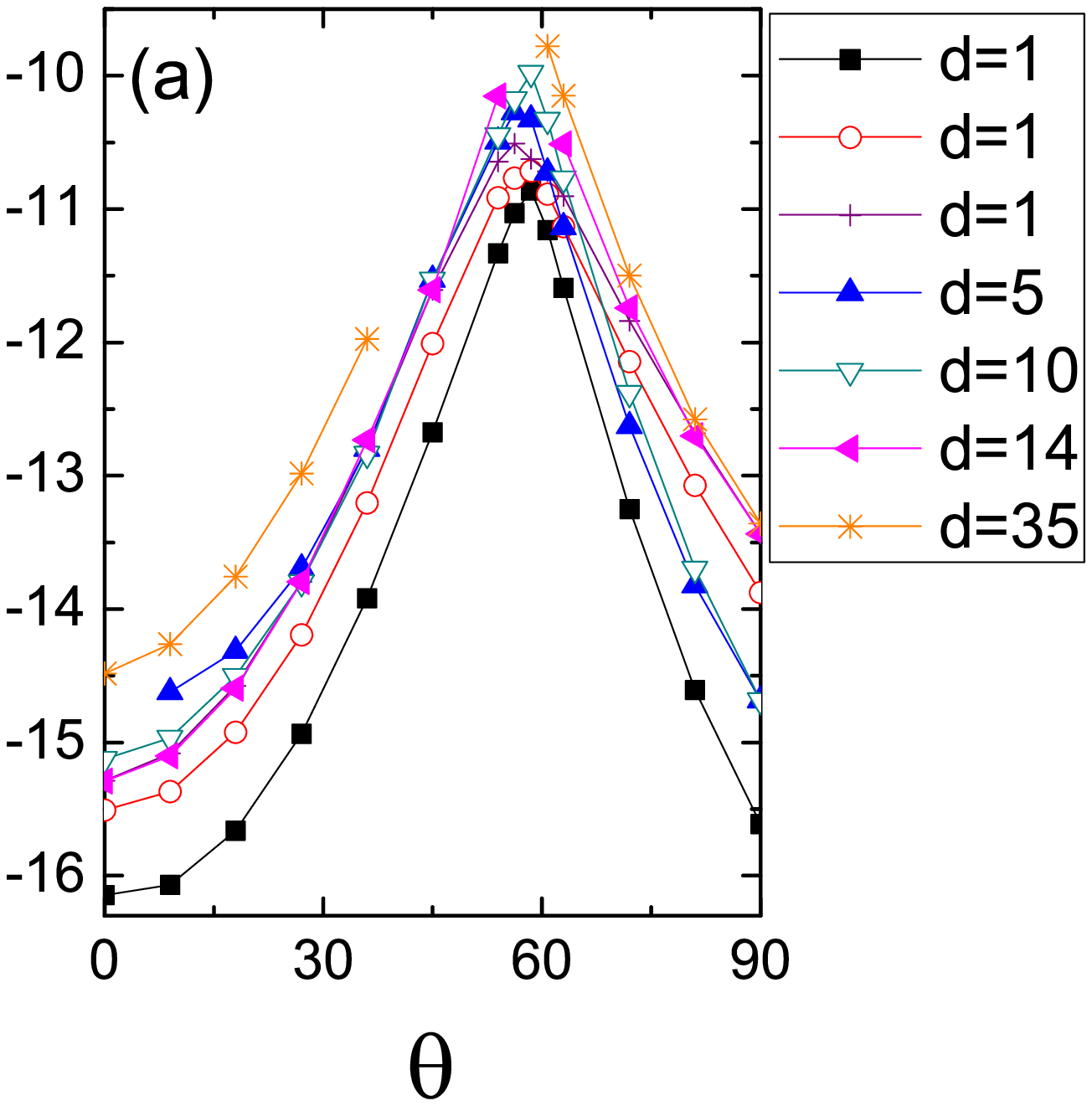,clip=1,width=0.55\linewidth,angle=0}
\centering\epsfig{file=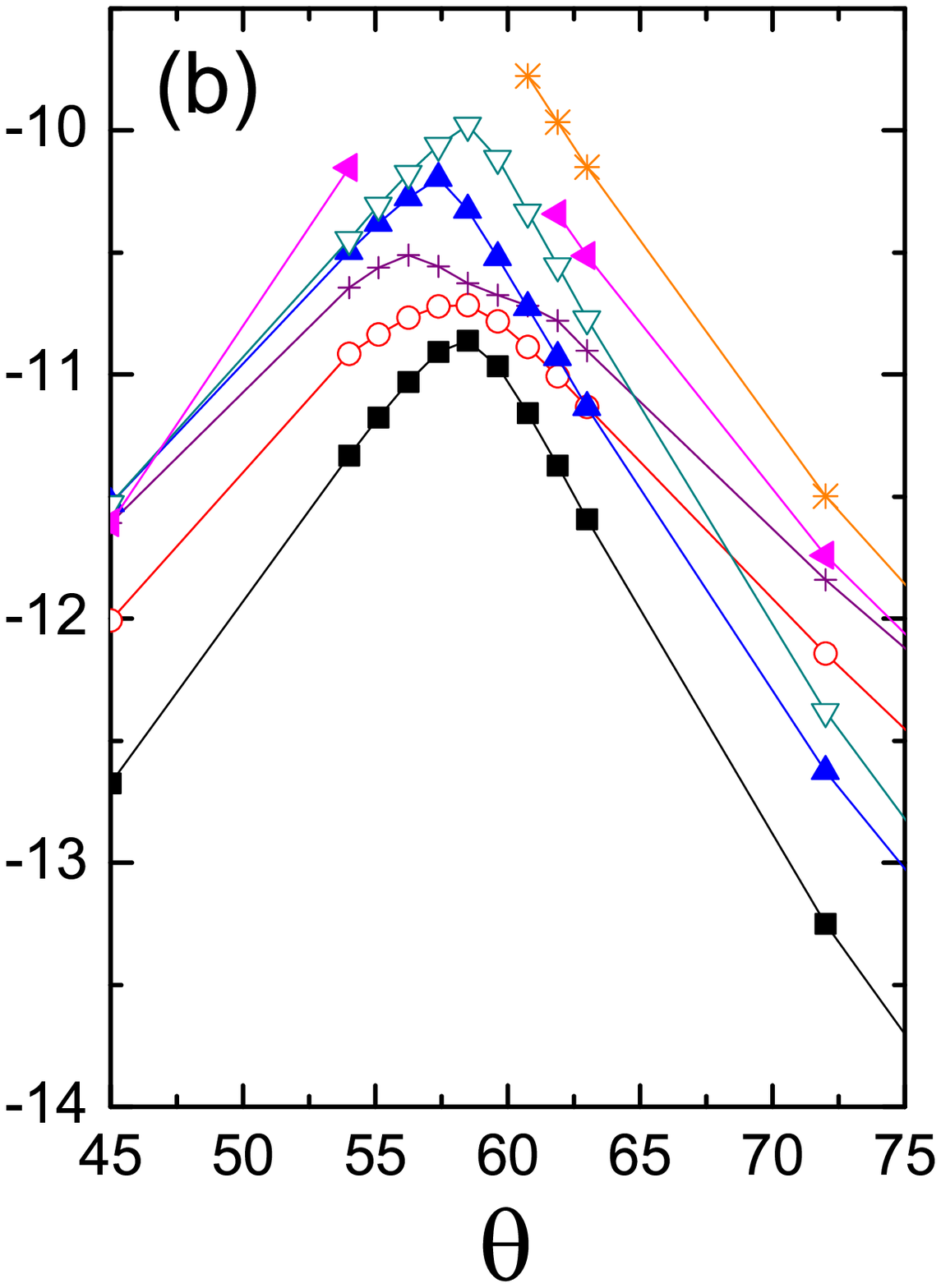,clip=1,width=0.41\linewidth,angle=0}
\caption{(a) The low-lying states for the $4\times 4$ cluster
at various values of $\theta$.
The dimensions of the corresponding $Sp(4)$ representations $d$
are marked.
The GS wavefunctions are always $Sp(4)$ singlets.
(b) The zooming-in around $\theta\approx 63^\circ$ exhibiting
various energy level crossings.
}
\label{fig:2Dspectra}
\end{figure}

The low-lying energy spectra for the $N=4 \times 4$ clusters
for $0<\theta<90^\circ$ are displayed in Fig. \ref{fig:2Dspectra}.
The ground states for all the values of $\theta$ are $Sp(4)$ singlets with
Casimir $C=0$, and that at $\theta=45^\circ$ is an $SU(4)$ singlet.
The lowest excited states are also $Sp(4)$ singlet states at
$\theta<63^\circ$.
The lowest spin multiplets appear as the 14-fold degenerate $Sp(4)$
symmetric tensor states with $C=10$.
A level crossing of the lowest excited states appears around
$\theta = 63^{\circ}$ implying that there exists competing phases nearby.
At $\theta>63^\circ$, the lowest excited states become 5-fold degenerate
$Sp(4)$ vector states with the Casimir $C=4$.
Another 10-fold degenerate states, which form the $Sp(4)$ adjoint
representation with $C=10$, appear as the next lowest excited states.
At the $SU(4)_B$ line, i.e., $\theta=90^\circ$, these two sectors
merge into the 15-fold degenerate states
forming the adjoint representation of the $SU(4)$ group whose
$SU(4)$ Casimir is $C=8$.

In Sec. \ref{sect:exact}, the appearance of the $Sp(4)$ singlet as
the lowest excited states in the small size systems implies the
dimerization in the thermodynamic limit.
This is confirmed in the large size DMRG results in Sec. \ref{sect:DMRG}.
Similarly, in the case of the $4\times 4$ cluster, the lowest excited
states are also $Sp(4)$ singlet at $\theta<63^\circ$.
This also suggests the spin disordered ground state with broken
translational symmetry in the thermodynamic limit.
Moreover, the gap between the GS and lowest
singlet excited state is very small in a narrow regime (roughly
$50^{\circ} \sim 60^{\circ}$), which implies that an intermediate
phase may exist exhibiting a different translational symmetry
breaking pattern from that with small values of $\theta$. However,
unlike the 1D case where we can justify the dimerization through
finite-size scaling of the vanishing of the $Sp(4)$ singlet-singlet
gap, it is impossible in 2D to detect the presence of the dimer
states or plaquette states from the exact diagonalization results.
Thus we will resort to other approaches to investigate the GS
profile in the following sections.

\begin{figure}[!htb]
\centering\epsfig{file=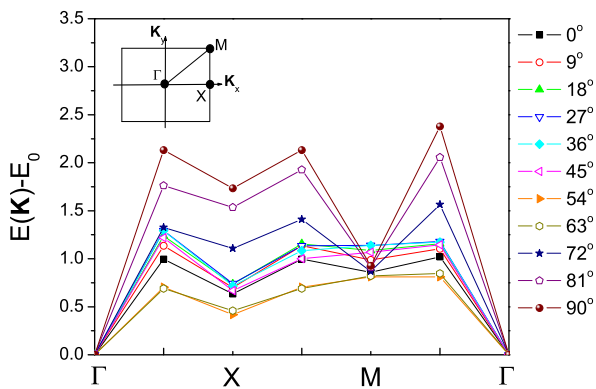,clip=1,width=0.9\linewidth,angle=0}
\caption{The energy dispersion $E(\vec K)-E_0$ v.s. $\theta$
for the $4 \times 4$ cluster.
$\Gamma$, M, X are the high symmetry points for the many-body
ground state momenta, corresponding to $\vec K=(0,0)$, $(\pi,\pi)$ and
$(\pi,0)$ respectively, in the first Brillouin zone.}
\label{fig:2Ddispersion}
\end{figure}

To further clarify, in Fig. \ref{fig:2Ddispersion} we present the
spectra of lowest energy states at each crystal momentum of
$\Gamma=(0,0)$, $X=(\pi, 0)$, and $M=(\pi, \pi)$, respectively.
At $\theta \le 63^{\circ}$, the states at the $X$-point are lower than
those at $M$-point, which are $Sp(4)$ singlets with the Casimir $C=0$.
These lowest singlet excitations along $(\pi,0)$ or $(0,\pi)$
would allow the GS to shift a lattice constant along $x$ or $y$-direction,
if the gap between these singlets vanishes in the thermodynamic limit.
It would implies a four-fold degeneracy in the thermodynamic limit
breaking translational symmetry.

In comparison, as $\theta \ge 72^{\circ}$, the energy of states at
the $M$-point are lower than those at the $X$-point, which are spin
multiplet with 10-fold degeneracy and the $Sp(4)$ Casimir $C=6$.
Actually, these states are not the lowest excited states which are
5-fold degenerate located at the $\Gamma$-point. Nevertheless, their
energy splitting from the 10-fold states is very small as shown in
Fig. \ref{fig:2Dspectra}. In the thermodynamic limit, inspired by
the QMC result of the occurrence of the long-range ordering in the
$SU_B(4)$ case, we infer the long-range staggered Neel ordering of
the $Sp(4)$ spin operators $L_{ab}$ and a long-range uniform
ordering of $Sp(4)$ vector operators $n_a$. Thus we infer a phase
transition from spin disordered ground state to the Neel-like state
breaking $Sp(4)$ symmetry.

Let us make an analogy with the 2D spin-$\frac{1}{2}$ $J_1$-$J_2$ model
\cite{dagotto1989a,dagotto1989b}.
In that case, the behavior of the low-lying energy levels indicates
that the lowest excited states with nonzero momentum are triplet
while the system is a magnetic Neel ($J_2/J_1 \lesssim 0.4$) and
collinear state ($J_2/J_1 \gtrsim 0.6$), corresponding to ${\bf
K}=(\pi,\pi)$ and $(\pi,0)$, respectively.
However, there exists an intermediate phase in $0.4 < J_2/J_1 < 0.6$,
where the GS is a magnetic disordered state and the lowest excited
state with nonzero momentum, $\vec K=(\pi,0)$, is singlet.
In this region it has been conjectured that the GS is a dimerization
state or a spin-liquid (resonated-valence-bond state).
Similarly, the low-lying energy behavior in our model implies
that the GS is non-magnetic at $\theta < 63^{\circ}$.
On the other hand, at $\theta \ge 63^{\circ}$, the GS has spinful
excitations and is relevant to the Neel state.

\subsection{The magnetic structure form factor}
\label{sect:mag}

\begin{figure}[!htb]
\centering\epsfig{file=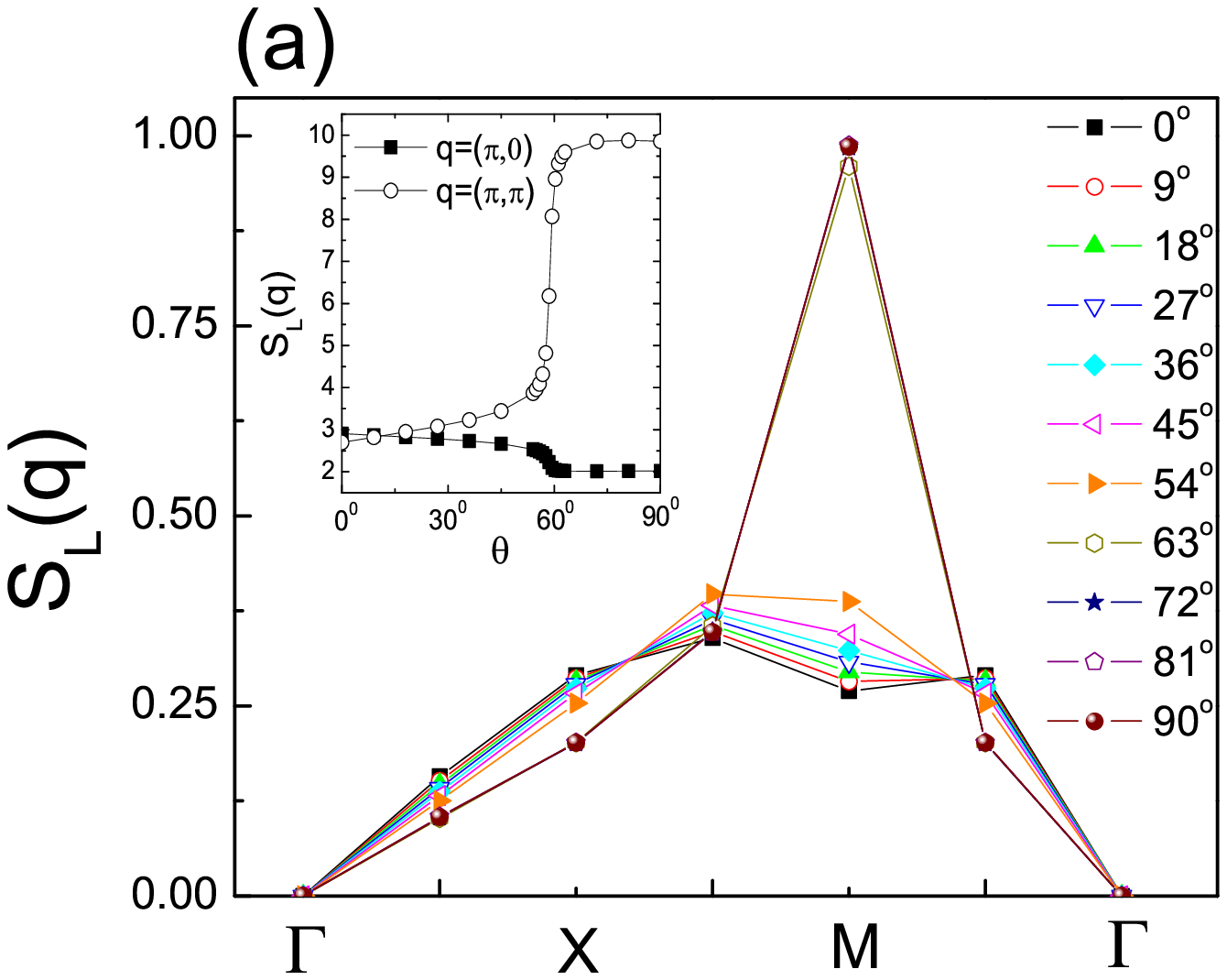,clip=1,width=0.9\linewidth,angle=0}
\centering\epsfig{file=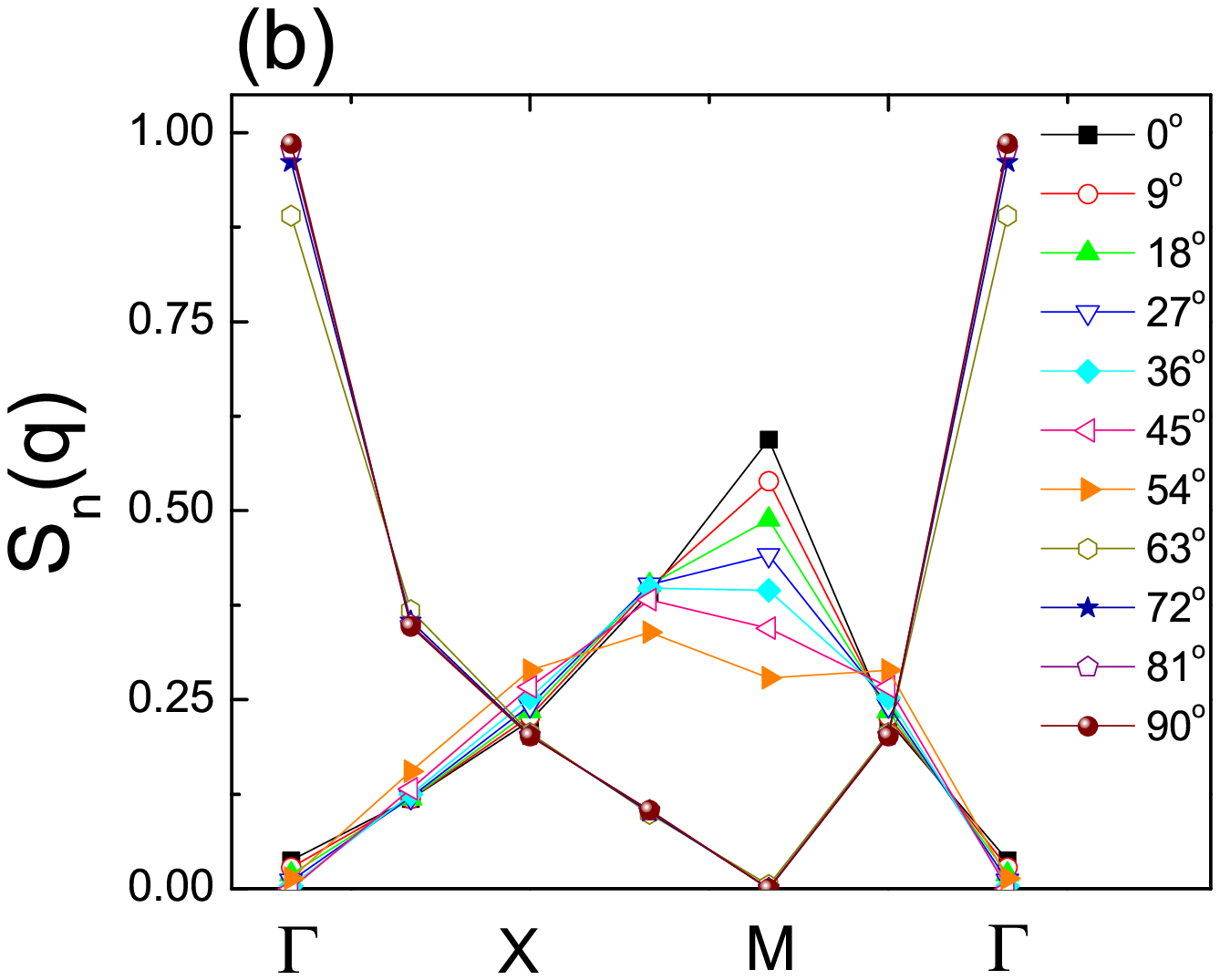,clip=1,width=0.9\linewidth,angle=0}
\caption{ The magnetic structure factors for the $4
\times 4$ cluster.
(a) The structure factors $S_L(\vec q)$ for the $Sp(4)$ generator sector
The inset is the comparison between $S_L(\pi,0)$
and $S_L(\pi,\pi)$ versus $\theta$.
(b) The $Sp(4)$ vector structure factor $S_n(\vec{q})$.
} \label{fig:neel}
\end{figure}

In this subsection, we present the results of the magnetic structure
form factors for the $N=4\times 4$ cluster. Two different structure
form factors $S_L(\vec q)$ and $S_n(\vec q)$ are defined for the
$Sp(4)$ generator and vector channels, respectively, as \bea
\label{eq:stucr} S_L(\vec{q})&=&\frac{1}{g_L N^2}\sum_{i,j, 1\le
a<b\le5} e^{i\vec{q} \cdot (\vec{r}_i-\vec{r}_j)}
\langle G| L_{ab}(i) L_{ab}(j)|G \rangle, \nn \\
S_n(\vec{q})&=&\frac{1}{g_n N^2}\sum_{i,j, a=1\sim 5} e^{i\vec{q}
\cdot (\vec{r}_i-\vec{r}_j)}\langle G|  n_a(i) n_a(j) |G\rangle,
\eea where the normalization constants $g_L=10$ and $g_n=5$.
$S_L(\vec q)$ and  $S_n(\vec q)$ are the analogy of the Fourier
transformation of $\langle G| \vec{S}_i \cdot \vec{S}_j | G \rangle$
in $SU(2)$ systems. If the long-range magnetic order appears, the
magnetic structure factor converges to a finite value in the
thermodynamic limit \cite{harada2003,schulz1992}.

The ED results of $S_L(\vec{q})$  for the $4 \times 4$ cluster
are presented in Fig. \ref{fig:neel} (a).
As $\theta\lesssim 60^\circ$,
$S_L(\vec q)$ distributes smoothly over all the momenta, and its
maximum is located at $\vec q =(\pi, \frac{\pi}{2})$, which is
slightly larger than other values of $\vec q$. In contrast, when
$60^\circ\lesssim\theta\le 90^\circ$, $S_L(\vec q)$ peaks at $\vec
K_M=(\pi,\pi)$.
The $Sp(4)$ vector channel structure factor $S_n(\vec q)$ is depicted in
Fig. \ref{fig:neel} (b).
At small values of $\theta$, it peaks at the M-point exhibiting
a dominate correlation at the momentum $(\pi,\pi)$.
As $\theta \gtrsim 60^\circ$, the peak changes to the $\Gamma$
point and the $M$-point becomes a minimum.

Along the $SU(4)_B$ line with $\theta=90^\circ$,
$S_n (\vec q)=S_L (\vec q + \vec K_M)$
due to the staggered definition of $Sp(4)$ vectors $n_a$
in Eq. \ref{eq:SU4B}.
This relation between $S_n(\vec{q})$ and $S_L(\vec{q}+\vec K_M)$
is consistent with the previous observation on the low-energy
spectra in Fig. \ref{fig:2Dspectra}. As $\theta \ge 60^{\circ}$
there are two nearly degenerate excited states beyond the GS, having
the total momentum of $(0,0)$ and $(\pi,\pi)$.
They correspond to the 5d vector representation with $C=4$ and the
10d tensor representation with $C=6$ in the $Sp(4)$ symmetry,
respectively.
The contributions to $S_{n}(\vec K_\Gamma)$ and $S_L(\vec K_M)$  mainly
come from the matrix
elements between the ground state and the 5d vector states,
and 10d antisymmetric tensor states, respectively.
On the other hand, in the case of $SU(4)_A$ with
$\theta=45^{\circ}$, $S_n(\vec{q})=S_L(\vec{q})$ for each $\vec{q}$.

These features highlight that the dominant Neel correlation of the
$Sp(4)$ generators $L_{ab}$'s not only exhibits along the $SU(4)_B$
line but also extends to a finite regime with non-zero values of
$J_2$. In the same parameter regime, the $Sp(4)$ vectors $n_a$'s
exhibit dominant uniform correlations. The critical value of
$\theta$ of the onset of the outstanding $S_L(\pi,\pi)$ is in good
agreement with the location of the level crossing shown in Fig.
\ref{fig:2Dspectra}, implying a transition of the GS from a non-Neel
state to a Neel type. The inset in Fig. \ref{fig:neel} (a) compares
the $S_L(\vec{q})$ behavior at $\vec{q}=(\pi,0)$ and $(\pi,\pi)$
versus $\theta$. $S_L(\pi,0)$ changes little as varying $\theta$.
Therefore, it is inferred that only the Neel-type order exists at
$\theta$ close to 90$^\circ$. The magnetic ordering at $(\pi,0)$
should not appear in the 2D $Sp(4)$ system.

Next one may raise a natural question: what is the spin pattern for
the Neel-order state as $\theta \rightarrow 90^\circ$?
According to Eq. \ref{eq:SU4B}, its classic energy can be minimized
by choosing a staggered configuration for
$\avg{G|L_{15}(i)|G}=\avg{G|L_{23}(i)|G}=(-)^i \frac{1}{2}$
and a uniform configuration of $\avg{G|n_4(i)|G}=\pm\frac{1}{2}$.
These correspond to the staggered arrangement
in the 2D lattice by using the two components of
$F_z=\pm \frac{3}{2}$,  or by using the other two components of
$F_z=\pm \frac{1}{2}$.
These different classic Neel states can be connected by an
$Sp(4)$ rotation.

\subsection{The columnar dimer correlations}
\label{sect:dimer}

\begin{figure}[!htb]
\centering\epsfig{file=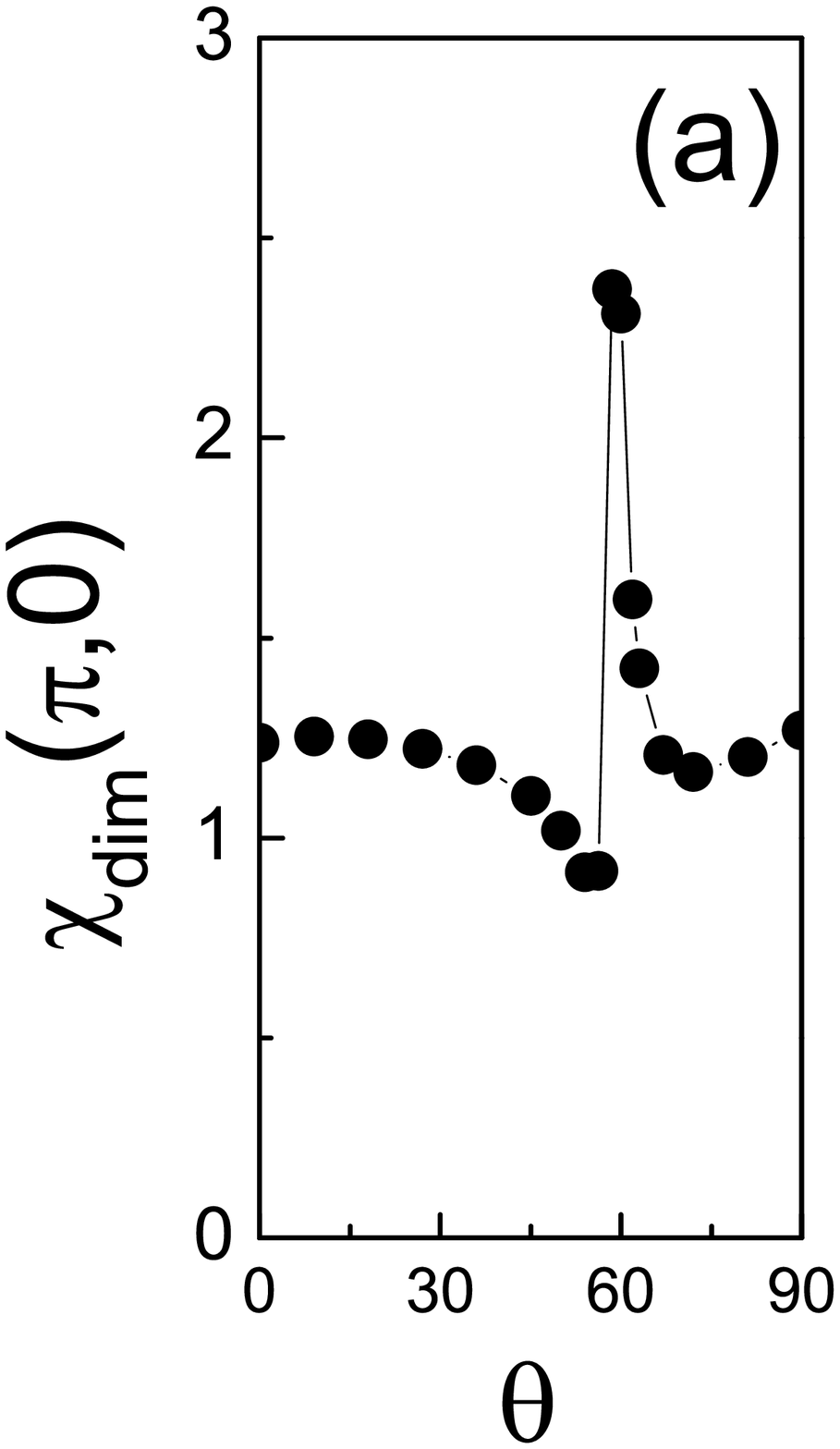,clip=1,width=0.31\linewidth,angle=0}
\centering\epsfig{file=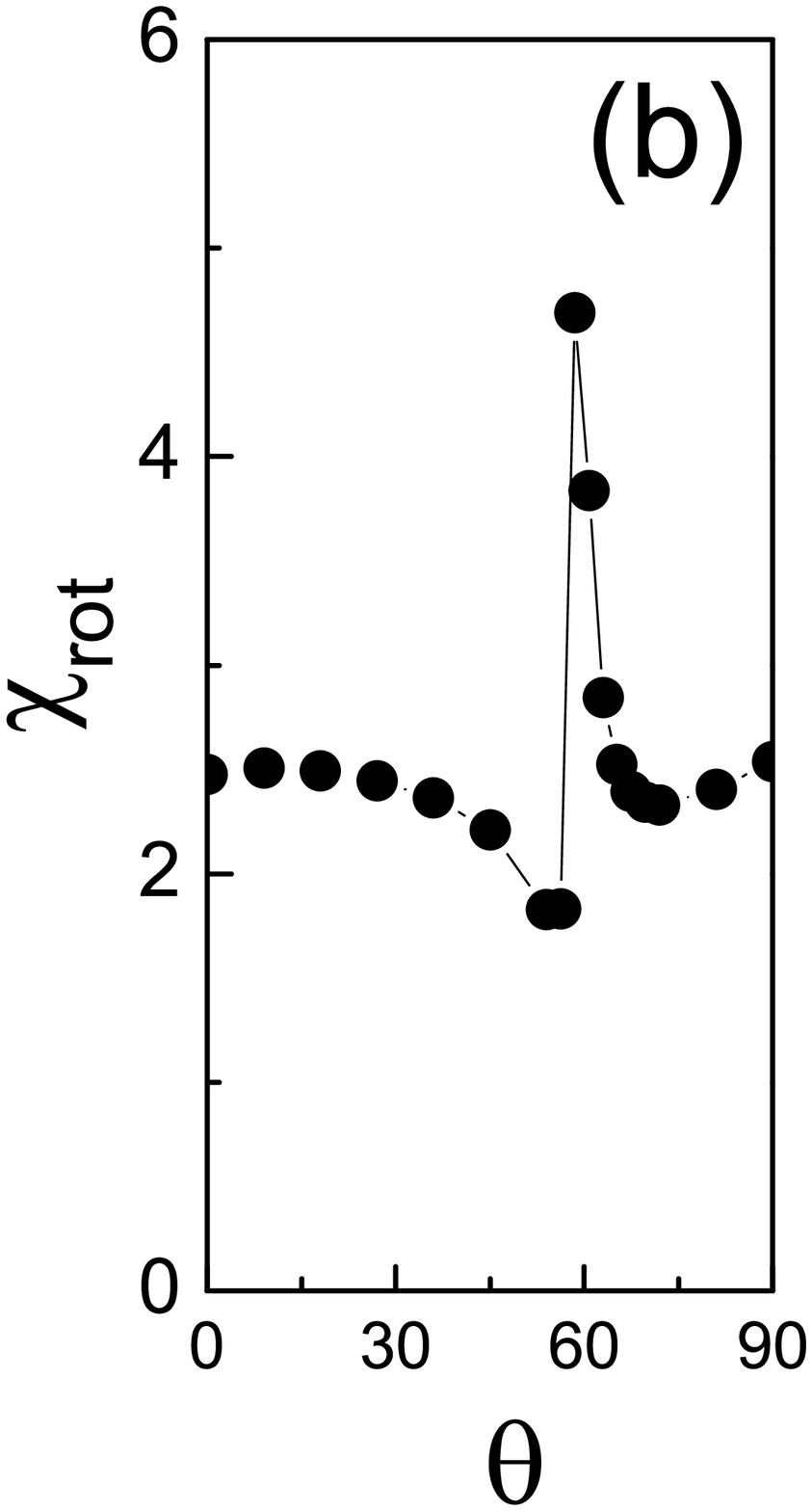,clip=1,width=0.285\linewidth,angle=0}
\centering\epsfig{file=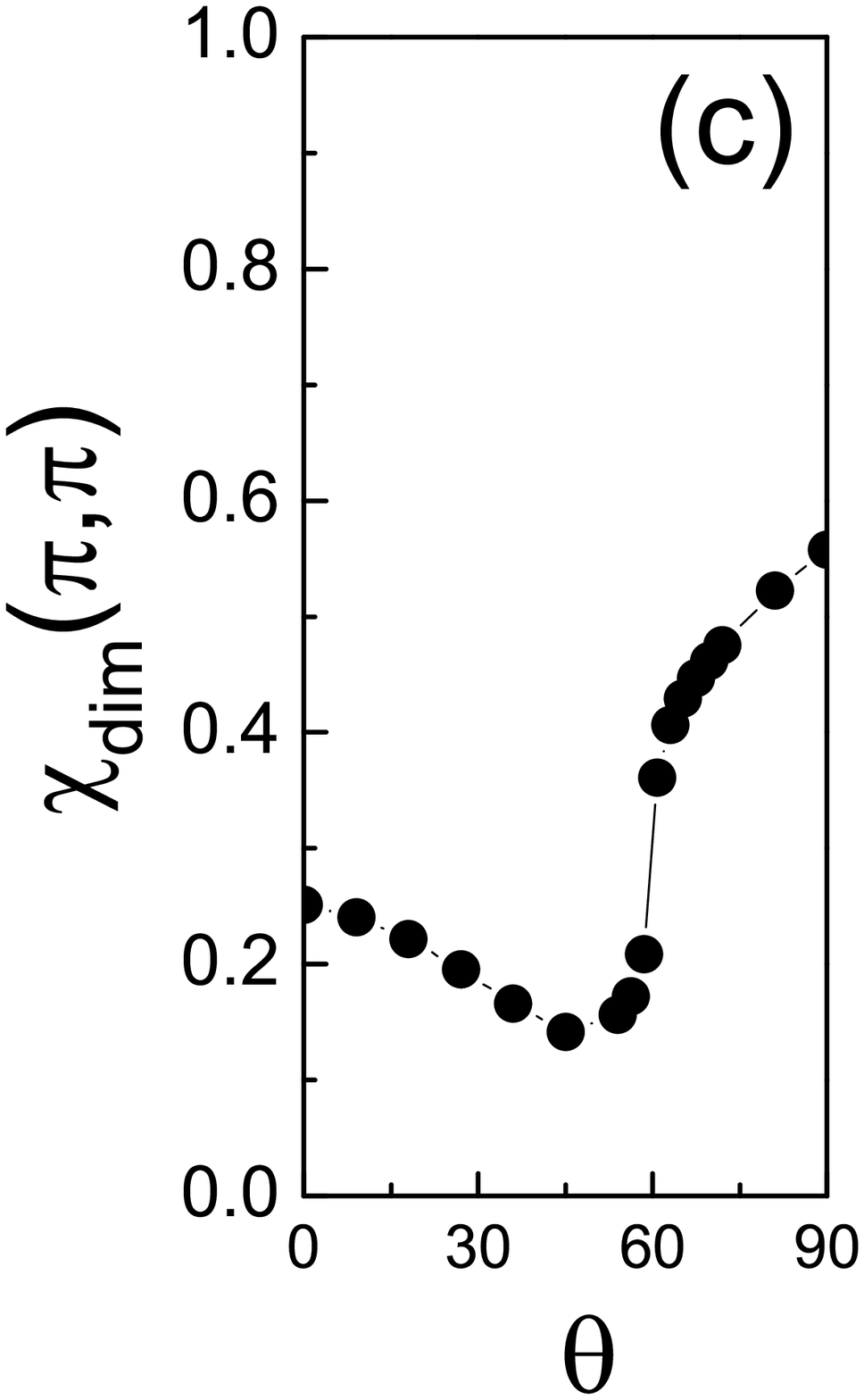,clip=1,width=0.33\linewidth,angle=0}
\caption{The susceptibilities defined in Eq. \ref{eq:sus} with
respect to the perturbations $O_{dim}$ and $O_{rot}$ for the
$N=4\times 4$ cluster. (a) $\chi_{dim}(\vec Q)$ versus $\theta$ at
$\vec Q=(\pi,0)$; (b) $\chi_{rot}$ versus $\theta$; (c)
$\chi_{dim}(\vec Q)$ versus $\theta$ at $\vec Q=(\pi,\pi)$. In both
cases, a small value of $\delta=0.01$ is taken to evaluate the
susceptibilities. Both susceptibilities exhibit peaks around
$\theta\approx 60^{\circ} \sim 70^{\circ}$.} \label{fig:2Dpert}
\end{figure}

In this subsection, we discuss the possibility of the dimer-ordered state
at intermediate values of $\theta$.
We define the susceptibility to a symmetry breaking perturbation as
\bea
\chi(\delta)=-\frac{2 [e(\delta)-e(0)]}{\delta^2},
\label{eq:sus}
\eea
where $e(0)$ is the GS energy per site given by the
Hamiltonian Eq. (\ref{eq:so5}) and $e(\delta)$ is
Eq. \ref{eq:so5} plus the corresponding perturbation
term $-\delta \hat{O}$\cite{santoro1999,capriotti2000}.
In the presence of long-range ordering, the corresponding susceptibility
$\chi=\lim_{\delta\to 0}\chi(\delta)$ will diverge in the thermodynamic limit.
It has been demonstrated that this approach can efficiently distinguish
dimerized and non-dimerized phases in the 1D $J_1$-$J_2$
spin-$\frac{1}{2}$ chain \cite{capriotti2000}, in which the phase
boundary $J_2/J_1 \approx 0.24$ between these two phases.

Here we employ the same method to study the dimerization correlations.
Although with small size calculations, we are unable to determine
the existence of long range order, it is still instructive
to observe the feature of $\chi$.
We have used it to test the 1D $Sp(4)$ system with the perturbation
term of $\hat{O}=\sum_i (-1)^i H_{ex}(i,i+1)$.
At $\theta = 60^{\circ}$, we found the dramatic growing behavior of
$\chi(\delta)$ upon decreasing $\delta$ and increasing the system size,
which leads to a divergent $\chi$ in the thermodynamic limit.
On the other hand, $\chi(\delta)$ at $\theta = 30^{\circ}$
has no tendency of divergence over decreasing $\delta$.
This observation is consistent with our previous analytical and numerical
studies: the 1D $Sp(4)$ system is either a gapless uniform liquid as
$\theta \le 45^{\circ}$ or a gapped dimerized state with
the breaking of translation symmetry at $\theta > 45^{\circ}$.

Next we apply this method to the 2D system with the size of $4 \times 4$,
and define two susceptibilities $\chi_{dim}(\vec Q)$ and $\chi_{rot}$
for two perturbations of $\hat O_{dim}(\vec Q)$ and $\hat O_{rot}$ as
\bea
\label{eq:perturbation1}
\hat{O}_{dim} (\vec Q)&=&\sum_{i} \cos (\vec{Q} \cdot \vec{r}_i)
H_{ex}(i,i+\hat{x}),  \\
\label{eq:perturbation2} \hat{O}_{rot}&=&\sum_{i}
[H_{ex}(i,i+\hat{x})-H_{ex}(i,i+\hat{y})], \eea where $H_{ex}(i,j)$
is defined as one bond of the Hamiltonian Eq. \ref{eq:so5} without
summation over $i$ and $j$. Let us set $\vec Q=(\pi, 0)$, thus
$\chi_{dim}(\pi,0)$ corresponds to the instability to the columnar
dimer configuration. Eq. \ref{eq:perturbation1} and Eq.
\ref{eq:perturbation2} break the translational symmetry along the
$x$-direction and rotational symmetry, respectively. The plaquette
ordering maintains the 4-fold rotational symmetry, thus will lead to
the divergence of $\chi_{dim}(\pi,0)$ but not $\chi_{rot}$. The ED
results for the susceptibilities with respect to the two
perturbations versus $\theta$ in Fig. \ref{fig:2Dpert} (a) and (b),
respectively. A small value of $\delta=0.01$ is taken. Both
susceptibilities exhibit a peak at $\theta$ from $60^\circ$ to
$70^\circ$, which implies a tendency to breaking both translational
and rotational symmetries. This shows that the columnar dimerization
instead of the plaquette ordering is a promising instability in this
regime in the thermodynamic limit. We have also calculated the
susceptibility of $\chi_{dim}(\vec Q)$ for $\vec Q=(\pi,\pi)$ which
corresponds to the instability to the staggered dimer configuration
in Fig. \ref{fig:2Dpert} (c). Although the magnitudes of
$\chi_{dim}(\pi,\pi)$ are smaller than $\chi_{dim}(\pi,0)$ and
$\chi_{rot}$, it suddenly raises up around $\theta=60^{\circ}$.

\subsection{The plaquette form factor}
\label{sect:plaquette}

\begin{figure}[!htb]
\centering\epsfig{file=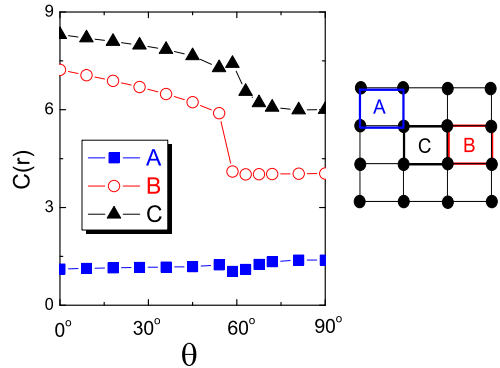,clip=1,width=0.75\linewidth,angle=0}
\caption{$C(\vec r)$ defined in Eq. \ref{eq:localspin} versus $\theta$.
The positions of plaquette $A$, $B$ and $C$ are defined on the right
schematics. $A$ (blue) is located at the corner whereas $C$
(black) in the most middle one.}
\label{fig:Fig5}
\end{figure}

In this subsection, we consider the plaquette type correlation. The
ground states of the $4\times 4$ cluster at $\theta\lesssim
60^\circ$ signal a different class from that in $\theta \ge
63^{\circ}$ in which the lowest excited states are spin multiplet
with $\vec K=(\pi,\pi)$. Here, the lowest excited states remain
$Sp(4)$ singlets with $\vec K=(\pi,0)$ or $(0,\pi)$. To further
elucidate the ground state profile, we define the local Casimir for
the plaquette centered at $\vec r$, \bea \label{eq:localspin} C(\vec
r) =\avg{G|\sum_{1\le a<b\le 5}\big\{\sum_{i} L_{ab}(i)\big\}^2 |G},
\eea where $i$ runs over the four sites of this plaquette. The
$SU(2)$ version of Eq. \ref{eq:localspin} has been used to classify
competing dimer and plaquette orders \cite{richter1996}. If the GS
exhibits a strong plaquette pattern, for instance, indicated as
phase {\bf A} in Fig. \ref{fig:2Dphasediagram}, the magnitudes of
$C(\vec r)$ will have obvious spatial variations between
nearest-neighboring plaquette. This is analogous to the 1D
dimerization picture in Fig. \ref{fig:nncorr} (a), where the
nearest-neighboring spin-spin correlations exhibit strong and weak
alternately in magnitude. When the spins around a plaquette are
strongly bound to form an $SU(4)$ singlet, $C(\vec r)$ should be
close to zero.

Fig. \ref{fig:Fig5} depicts the behavior of $C(\vec r)$ at various
values of  $\theta$ for the $4\times 4$ cluster. In order to
explicitly reflect the plaquette formation, we use {\it open}
boundary conditions rather than periodic boundary conditions. In
this case only $C_{4v}$ point group symmetry is applicable in the
ED. The $C(\vec r)$ for the corner plaquette $A$ is much smaller than
$\frac{1}{5}$ of  those at the center $C$ and the middle of the edge $B$ for
small values of $\theta$. This is in sharp contrast to the 2D
spin-$1/2$ model which renders $C(A)=0.545$, $C(B)=1.015$ and
$C(C)=1.282$, which only show the difference at order of 1. The
comparison suggests the pinning-down plaquette state in the 2D
$Sp(4)$ system under the open boundary.
We observe that $C(A)$ and $C(B)$ decrease while $\theta$ goes
beyond $60^{\circ}$.
It accounts for the formation of the plaquette-type pattern weakens or
even vanishes.

Combined the above observations, it is likely that for $\theta < 60^{\circ}$
the GS has a strong plaquette-like correlation, that could be the resonate
plaquette state proposed by Bossche {\it et al.}\cite{bossche2000}
or a certain spin-liquid. It does survive not
only along the $SU(4)_A$ line but also in a finite regime.
Nevertheless, we have to emphasize that this picture cannot be
conclusively determined due to finite size effects and further
larger size calculations are needed to confirm.


\section{Conclusion}
In conclusion, we study an $Sp(4)/SO(5)$ spin Heisenberg model which
can be realized by the large spin ultra-cold fermions with
$F=\frac{3}{2}$. The $Sp(4)$ Heisenberg model describing the
magnetic exchange at the insulating state of quarter-filling is
simulated by exact diagonalization and DMRG. In 1D, our numerical
results are in agreement with previous analytic studies. There are
two competing phases: a gapped dimer phase with spin gap at $\theta>
45^{\circ}$ and a gapless spin liquid at $\theta \le 45^{\circ}$.
The phase boundary is identified as $\theta=45^{\circ}$ which
belongs to $SU(4)_A$-type symmetry. In the gapless spin liquid
phase, the static correlation functions decay algebraically with
four-site periodicity oscillations.

We also investigate the $Sp(4)$ spin model on a 2D square lattice up
to $16$ sites by means of exact diagonalization methods. Our
numerical results show three competing correlations: Neel-type,
plaquette formation and columnar spin-Peierls dimerization,
depending on $\theta$'s. Such observation can have phase behavior
analogy in comparison with the speculated phase diagram depicted in
Fig. \ref{fig:2Dphasediagram}. Due to the finite size effects,
however, we are unable to conclusively identify the existence of
these phases and the phase boundaries based on the small cluster.
More numerical studies are necessary to further explore the phase
diagram in the thermodynamic limit.

\acknowledgements H. H. H. is grateful to helpful discussions with
Stephan Rachel and computational facilities from Tunghai University.
H. H. H. also appreciates Zi Cai and Cheng-Chien Chen for fruitful
discussions and suggestions on exact diagonalization techniques. H.
H. H. and C. W. are supported by NSF under No. DMR-0804775. Y. P. W.
is supported by NSFC and 973-project of MOST China.

\appendix
\section{Representation theory of the simple Lie groups and algebras}
The representation theory of Lie groups and algebras
can be found in standard group theory textbooks \cite{ma}.
Here we give a brief pedagogical introduction.
Among the group generators, we choose the maximal set
of generators that commute with each other as the {\it Cartan sub-algebra}
$\{H_i, (i=1,...k)\}$, where $k$ is called the rank of the Lie algebra.
For example, the $SU(2)$ algebra is rank one,
whose Cartan sub-algebra only contains $S_z$.
All other generators can be organized as eigen-operators of
each generator in the Cartan sub-algebra, which are called {\it roots}.
Roots always appear in terms of Hermitian conjugate pairs as
$E_{j\pm}$ with the relation $E_{j-}=E^\dagger_{j+}$.
They satisfy the commutation relations of
\bea
[H_i, E_{\pm j}]= \alpha_{\pm j}(i) ~E_{\pm j}, \ \ \,
\eea
with $\vec \alpha_{j}=-\vec \alpha_{-j}$,
where the $i$-th elements of the vectors $\vec \alpha_{\pm j}$
are the eigenvalue of $E_{\pm j}$ with respect to $H_i$.
For example, for the simplest $SU(2)$ case, the roots
are $S_\pm=S_x\pm i S_y$ and $[Sz, S_\pm]=\pm S_\pm$,
where $\alpha_\pm$ only have one component with $\alpha_{\pm}=\pm 1$.

Among all the roots, we fix the convention to use $E_{+j}$ for
{\it positive roots}, which means the first non-zero components
of their $\vec \alpha_{+j}$  are positive.
Positive roots can be decomposed into the linear combinations of
{\it simple roots} with non-negative integer coefficients.
The number of simple roots of a simple Lie algebra equals to its
rank.
The {\it Cartan matrix} $A$ of a simple Lie algebra is defined as
\bea
A_{ij}=2 \frac{(\vec \alpha_i,\vec \alpha_j)}{(\vec \alpha_i,\vec \alpha_i)}, ~~~
(i, j=1,...,k),
\eea
where $\vec \alpha_i$ is the vector of eigenvalues of the simple
root $E_i$; the inner products of $\alpha$-vectors is
defined as
\bea
(\vec \alpha_i, \vec \alpha_j)=\sum_{l=1}^k \alpha_i (l) \alpha_j( l).
\eea
The dimension of the Cartan matrix is the same as the Cartan sub-algebra.
For the $SU(2)$ group, the only positive and simple root is $S_+$,
and the $1\times 1$ Cartan matrix $A=2$.

An important concept of the representations of the simple Lie algebra
is {\it weight}.
For a rank-$k$ simple Lie algebra, its fundamental weights can be
solved through its $k\times k$ Cartan matrix
\bea
\vec {M}_i= \sum_j \vec{\alpha}_j (A^{-1})_{ji}, ~~~(i=1, ..., k).
\label{eq:fdwght}
\eea
Any irreducible representation of a simple Lie algebra is uniquely
determined by its highest weight $\vec M^*$, which can be written
as a linear combination of the fundamental weights $\vec M_i$
\bea
\vec M^*=\sum_i \mu_i \vec M_i, (i=1,..., k),
\eea
where $\mu$'s are non-negative integers.
The dimension of the representation $M^*$ is
\bea
d(M^*)&=& \prod_{\mbox{positive roots}} \Big[
1+ \frac{ (\vec M^* ,\vec \alpha_i)}{(\vec R, \vec \alpha_i)} \Big],
\label{eq:dim}
\eea
with
\bea
\vec R&=& \frac{1}{2} \sum_{\mbox {positive roots}} \vec \alpha_i.
\label{eq:R}
\eea
Please notice that the product in Eq. \ref{eq:dim} and summation
in Eq. \ref{eq:R} take over all the positive roots.
The value of the {\it Casimir} operator for the representation denoted by $M^*$
is
\bea
C (\vec M^*)= (\vec M^* , \vec M^* + 2 \vec R).
\eea
For the simplest example of $SU(2)$, the only fundamental weight
$M=\frac{1}{2}$.
The highest weights is just $M^*=S$, where $S$ takes half-integer
and integer numbers.
Obviously, $d(S)=2S+1$ and $C(S)=S(S+1)$ as expected.

\section{The $Sp(4) (SO(5))$ algebra}
For convenience, we use the following symbols to represent the
$Sp(4) (SO(5))$ generators $L_{ab} (1\le a <b \le 5)$ defined in Eq.
\ref{eq:generators} as \bea L_{ab}=\left( \begin{array}{ccccc}
0& \Re \pi_x & \Re \pi_y &\Re \pi_z& Q\\
 &   0       & -S_z      & S_y     & \Im \pi_x\\
 &           &  0        & -S_x    & \Im \pi_y \\
 &           &           & 0       & \Im \pi_z \\
 &           &           &         &   0
\end{array} \right),
\eea
where
\bea
&&\pi_x^\dagger=\Re \pi_x+i \Im \pi_x=
\psi^\dagger_{3\over2} \psi_{-{3\over2}}+\psi^\dagger_{1\over2} \psi_{-{1\over2}},
\nonumber \\
&&\pi_x= \Re \pi_x-i \Im \pi_x=
\psi^\dagger_{-3\over2} \psi_{{3\over2}}+\psi^\dagger_{-{1\over2}} \psi_{{1\over2}},
\nonumber \\
&&\pi_y^\dagger=\Re \pi_y+i \Im \pi_y=
-i(\psi^\dagger_{3\over2} \psi_{-{3\over2}}-\psi^\dagger_{1\over2} \psi_{-{1\over2}}),
\nonumber \\
&&\pi_y=\Re \pi_y-i \Im \pi_y=
i(\psi^\dagger_{-3\over2} \psi_{{3\over2}}-\psi^\dagger_{-{1\over2}} \psi_{{1\over2}}),
\nonumber\\
&&\pi_z^\dagger=\Re \pi_z+i \Im \pi_z=
\psi^\dagger_{3\over2} \psi_{-{1\over2}}-\psi^\dagger_{1\over2} \psi_{-{3\over2}},
\nonumber \\
&&\pi_z= \Re \pi_z-i \Im \pi_z=
\psi^\dagger_{-{1\over2}} \psi_{{3\over2}}-\psi^\dagger_{-{3\over2}} \psi_{{1\over2}},
\nonumber \\
&& S_+=S_x+iS_y=
\psi^\dagger_{3\over2} \psi_{1\over2}-\psi^\dagger_{-{1\over2}} \psi_{-{3\over2}},
\nonumber \\
&& S_-=S_x-iS_y=
\psi^\dagger_{1\over2} \psi_{3\over2}-\psi^\dagger_{-{3\over2}} \psi_{-{1\over2}},
\nonumber \\
&&
S_z ={1\over 2} ( \psi^\dagger_{3\over2} \psi_{{3\over2}}
-\psi^\dagger_{1\over 2} \psi_{{1\over2}}+
\psi^\dagger_{-{1\over2}} \psi_{-{1\over2}}- \psi^\dagger_{-{3\over2}}
\psi_{-{3\over2}}), \nonumber \\
&&Q ={1\over 2} ( \psi^\dagger_{3\over2} \psi_{{3\over2}}
+\psi^\dagger_{1\over 2} \psi_{{1\over2}}-
\psi^\dagger_{-{1\over2}} \psi_{-{1\over2}}-\psi^\dagger_{-{3\over2}}
\psi_{-{3\over2}}). \nn \\
\eea

\begin{table}[h]\label{so5}
\begin{center}
\begin{tabular}{|c|c|}   \hline
($Q, S_z$) & Roots  \\ \hline
$\alpha_{\pm1}=\pm(1,-1)$&
$E_1= {1\over  \sqrt{24}} (\pi_x^\dagger -i\pi_y^\dagger);
E_{-1}= {1\over  \sqrt{24}} (\pi_x +i\pi_y)$ \\ \hline
$\alpha_{\pm2}=\pm(0,1)$&
$E_2= {1\over  \sqrt{12}} (S_x +i S_y);
E_{-2}= {1\over \sqrt{12}} (S_x -i S_y)$ \\ \hline
$\alpha_{\pm3}=\pm(1,1)$&
$E_3= {1\over  \sqrt{24}} (\pi_x^\dagger +i\pi_y^\dagger);
E_{-3}= {1\over \sqrt{24}} (\pi_x -i\pi_y)$ \\ \hline
$\alpha_{\pm4}=\pm(1,0)$&
$E_4= {1\over   \sqrt{12}}  \pi_z^\dagger;
E_{-4}= {1\over \sqrt{12}} \pi_z$ \\ \hline
\end{tabular}
\caption{Cartan sub-algebra and its roots.
$[E_1,E_{-1}]= \frac{1}{6} (Q-S_z)$,
$[E_2,E_{-2}]= \frac{1}{6} S_z$,
$[E_3,E_{-3}]= \frac{1}{6} (Q+S_z)$,
$[E_4,E_{-4}]= \frac{1}{6} Q$.}
\end{center}
\end{table}

\begin{table}[h]\label{table:so5}
\begin{center}
\begin{tabular}{|c|c|c|c|c|c|}   \hline
&($\mu_1,\mu_2$) & $M^*$& $d(M^*)$ & $C(\vec M^*)$   \\ \hline
1&(0,0) & (0,0)& 1&0\\ \hline
2&(0,1) & ($\frac{1}{2},\frac{1}{2}$) & 4& $\frac{5}{2}$ \\ \hline
3&(1,0) & (1,0) &5&4 \\ \hline
4&(0,2) & (1,1) &10&6  \\ \hline
5&(2,0) & (2,0) &14&10\\ \hline
6&(1,1) & ($\frac{3}{2},\frac{1}{2}$)&16&$\frac{15}{2}$  \\ \hline
7&(1,2) & (2,1)&35&12  \\ \hline
8&(0,3) & $(\frac{3}{2},\frac{3}{2})$& 20 & $\frac{21}{2}$  \\ \hline
\end{tabular}
\caption{Some irreducible representations of the $Sp(4)/SO(5)$
group: the highest weights, dimensions and Casimirs.}
\end{center}
\end{table}

The ten operators of $Sp(4)$ satisfy the commutation relations \bea
[L_{ab}, L_{cd}]=-i (\delta_{bc}L_{ad} +\delta_{ad} L_{bc}
-\delta_{ac}L_{bd}-\delta_{bd} L_{ac}), \ \ \ \eea which is rank-2
Lie algebra. Its Cartan sub-algebra only contains two commutable
generators $H_i (i=1,2)$, which can be chosen as $(H_1=S_z, H_2=Q)$.
We group the other 8 generators as their eigen-operators, {\it
i.e.}, roots as represented $E_{\pm 1}, E_{\pm 2}, E_{\pm 3}, E_{\pm
4}$, whose eigenvalue vectors $\vec \alpha_{\pm j}$ are presented in
Tab. I. The simple roots are $E_1$ with $\vec \alpha_1=(1,-1)$ and
and $E_2$ with $\vec \alpha_2=(0,1)$. The other roots can be
represented as $E_3=E_1+2 E_2, E_4=E_1+E_2$. The $Sp(4)/SO(5)$
Cartan matrix reads \bea A =\left( \begin{array}{cc} 2&-1\\ -2&2
\end{array} \right). \eea

We solve the fundamental weights
as $\vec M_1=(1,0), \vec M_2=(\frac{1}{2},\frac{1}{2})$
from Eq. \ref{eq:fdwght}.
The highest weight $\vec M^*$ can be written as
\bea
\vec M^*= (m_1,m_2)=(\mu_1+ \frac{\mu_2}{2}, \frac{\mu_2}{2}),
\eea
where $\mu_{1,2}$ are non-negative integers.
The dimension of the corresponding representation is
\bea
d(\mu_1,\mu_2)&=&(1+\mu_1) (1+\mu_2) (1+ {2\mu_1+\mu_2 \over 3})
\nn \\
&\times& (1+{\mu_1+\mu_2 \over 2}).
\eea
The representation $(\mu_1,\mu_2)$ belongs to the category of tensor
or spinor representations of $Sp(4)$ when $\mu_2$ is even or odd,
respectively.
The Casimir operator reads
\bea
C(\vec M^*)&=&\sum_{a<b} L_{ab}^2=Q^2+S_z^2
+6\sum_{\alpha} \{E_\alpha, E_{-\alpha}\}
\nonumber \\
&=&m_1(m_1+3)+m_2(m_2+1).
\eea

We summarize some frequently used representations of $Sp(4)(SO(5))$
in Table II.
The representations with indices $1$ to $5$ are particularly useful.
They are the identity (1d),
the fundamental spinor (4d), vector (5d), adjoint (10d), symmetric traceless
tensor (14d) representations of the $SO(5)$ group, respectively.

\section{$SU(4)(SO(6))$ algebra}

The $SU(4)$ group is isomorphic to $SO(6)$. Their relation is
similar to that between $SU(2)$ and $SO(3)$, or $Sp(4)$ and $SO(5)$.
As represented in Eq. \ref{eq:generators}, $L_{ab}$ and the five
spin-quadrapole operators $n_a=\psi^\dagger_\alpha
\Gamma_{\alpha\beta}^a \psi_\beta$ together form the 15 generators
of the $SU(4)$ group. Explicitly, the generators of $n_a$s are
written as \bea n_1 &=&{i\over 2} ( \psi^\dagger_{3\over2}
\psi_{-{1\over2}} +\psi^\dagger_{1\over 2} \psi_{-{3\over2}}-
\psi^\dagger_{-{1\over2}} \psi_{3\over2}- \psi^\dagger_{-{3\over2}}
\psi_{1\over2}), \nonumber \\
n_2 &=&{1\over 2} ( \psi^\dagger_{3\over2} \psi_{{1\over2}}
+\psi^\dagger_{1\over 2} \psi_{{3\over2}}-
\psi^\dagger_{-{1\over2}} \psi_{-{3\over2}}- \psi^\dagger_{-{3\over2}}
\psi_{-{1\over2}}) \nonumber \\
n_3 &=&-{i\over 2} ( \psi^\dagger_{3\over2} \psi_{{1\over2}}
-\psi^\dagger_{1\over 2} \psi_{{3\over2}}-
\psi^\dagger_{-{1\over2}} \psi_{-{3\over2}}+ \psi^\dagger_{-{3\over2}}
\psi_{-{1\over2}}) \nonumber \\
n_4 &=&{1\over 2} ( \psi^\dagger_{3\over2} \psi_{{3\over2}}
-\psi^\dagger_{1\over 2} \psi_{{1\over2}}-
\psi^\dagger_{-{1\over2}} \psi_{-{1\over2}}+ \psi^\dagger_{-{3\over2}}
\psi_{-{3\over2}}) \nonumber \\
n_5 &=&-{1\over 2} ( \psi^\dagger_{3\over2} \psi_{-{1\over2}}
+\psi^\dagger_{1\over 2} \psi_{-{3\over2}}+
\psi^\dagger_{-{1\over2}} \psi_{3\over2}+ \psi^\dagger_{-{3\over2}}
\psi_{1\over2}). \nn \\
\eea
The rank of the $SU(4)$ group is three.
We choose $(Q, S_z, n_4)$ as Cartan sub-algebra and group the other 12
generators as roots as shown in Tab. III.

\begin{table}[h]
\begin{center}
\begin{tabular}{|c|c|}   \hline
($Q, S_z,n_4$) & roots  \\ \hline
$\alpha_{\pm1}=\pm(1,-1,0)$&
$F_1={1\over \sqrt{32}} (\pi^\dagger_x-i\pi_y^\dagger)=
{1\over\sqrt{8}} \psi^\dagger_{1\over2} \psi_{-{1\over2}}$ \\ \hline
$\alpha_{\pm2}=\pm(0,1,-1)$&
$F_2={ (S_x +i S_y)-(n_2+i n_3)\over \sqrt{32}} =
 {1\over \sqrt{8}} \psi^\dagger_{-{1\over2}} \psi_{-{3\over 2}}$ \\ \hline
$\alpha_{\pm3}=\pm(0,1,1)$&
$F_3= {1\over  \sqrt{32}} (S_x +i S_y+n_2+i n_3) =
{1\over \sqrt{8}} \psi^\dagger_{3\over2} \psi_{1\over 2}$ \\ \hline
$\alpha_{\pm4}=\pm(1,1,0)$&
$F_4= {1\over  \sqrt{32}} (\pi_x^\dagger+i\pi_y^\dagger)
={1\over\sqrt{8}} \psi^\dagger_{3\over2} \psi_{-{3\over2}}
$ \\ \hline
$\alpha_{\pm5}=\pm(1,0,1)$&
$F_5= {1\over   \sqrt{32}}  (\pi_z^\dagger-i(n_1-in_5))
={1\over \sqrt{8}} \psi^\dagger_{3\over 2} \psi_{-{1\over2}}$ \\ \hline
$\alpha_{\pm6}=\pm(1,0,-1)$&
$F_6= {1\over   \sqrt{32}}  (\pi_z^\dagger+i(n_1-in_5))
={-1\over \sqrt{8}} \psi^\dagger_{1\over 2} \psi_{-{3\over2}}$ \\ \hline
\end{tabular}
\caption{Cartan sub-algebra and its roots.
$[F_1,F_{-1}]= \frac{1}{8}(Q-S_z),
[F_2, F_{-2}]= \frac{1}{8}(S_z-n_4),
[F_3, F_{-3}]= \frac{1}{8}(S_z+n_4),
[F_4, F_{-4}]= \frac{1}{8}(Q+S_z),
[F_5, F_{-5}]= \frac{1}{8}(Q+n_4),
[F_6, F_{-6}]= \frac{1}{8}(Q-n_4).$
 }
\end{center}
\end{table}

The simple roots are $F_1, F_2$ and $F_3$ with
the eigenvalue vectors $\vec \alpha_1=(1,-1,0)$,
$\vec \alpha_2=(0,1,-1)$, $\vec \alpha_3=(0,1,1)$, respectively.
The other positive roots are represented as $F_4=F_1+F_2+F_3, F_5=F_1+F_3$
and $F_6=F_1+F_2$.
The $SU(4)/SO(6)$ Cartan matrix reads
\bea
A
=\left( \begin{array}{ccc}
2&-1&-1\\-1&2&0\\-1&0&2 \end{array}
\right).
\eea
The fundamental weights can be solved by using Eq. \ref{eq:fdwght}
as
\bea
\vec M_1=(1,0,0), \vec M_2=(\frac{1}{2},\frac{1}{2},-\frac{1}{2}),
\vec M_3=(\frac{1}{2},\frac{1}{2},\frac{1}{2}).
\eea
The highest weight $\vec M^*$ of each representation can be chosen as
\bea
\vec M^*&=&(m_1,m_2,m_3)=\mu_1 \vec M_1+\mu_2 \vec M_2+\mu_3 \vec M_3 \nn \\
&=& (\mu_1+\frac{\mu_2}{2}+\frac{\mu_3}{2}, \frac{\mu_2}{2}+\frac{\mu_3}{2},
-\frac{\mu_2}{2}+\frac{\mu_3}{2}).
\eea
The dimension and Casimir of the representation $\vec M^*$ are
represented as
\bea
d(\vec M^*)&=&(1+\mu_1) (1+\mu_1) (1+\mu_1)(1+{\mu_1+\mu_2\over 2})  \nonumber \\
&\times&
(1+{\mu_1+\mu_3\over 2}) (1+{\mu_1+\mu_2+\mu_3\over 3}),\\
C(\vec M^*)&=&H_1^2+H_2^2+H_3^2  +8 \sum_{\Delta^+}
\big\{  F_\alpha, F_{-\alpha} \big\} \nonumber \\
&=& m_1 (m_1+4) +m_2 (m_2+2) +m_3^2.
\eea

\begin{table}[h]
\begin{center}
\begin{tabular}{|c|c|c|c|c|c|}   \hline
Rep&($\mu_1,\mu_2,\mu_3$) & $M^*(m_1,m_2,m_3)$& $d(M^*)$ & Casimir   \\ \hline
1&(0,0,0) & (0,0,0)  & 1&0\\ \hline
2&(0,0,1) & (${1\over2},{1\over2},{1\over2}$ ) & 4& $\frac{15}{4}$\\ \hline
3&(0,1,0) & (${1\over2},{1\over2},{-1\over2}$ )& 4& $\frac{15}{4}$  \\ \hline
4&(1,0,0) & (1,0,0) &6 &5\\ \hline
5&(0,0,2) & (1,1,1)  &10&9 \\ \hline
6&(0,2,0) & (1,1,-1) &10&9 \\ \hline
7&(0,1,1) & (1,1,0)  &15&8\\ \hline
8&(2,0,0) & (2,0,0)  &20&12 \\ \hline
\end{tabular}
\caption{Some frequently used irreducible representations of the
$SO(6)$ or $SU(4)$ group: the highest weight, dimension and
Casimir.}
\end{center}
\end{table}

\begin{figure}
\centering\epsfig{file=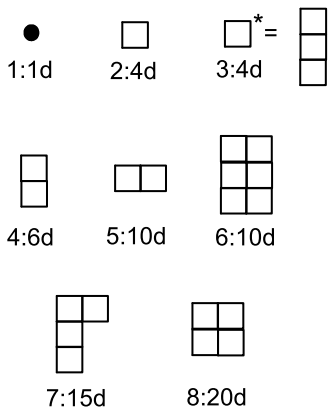,clip=1,width=0.5\linewidth,angle=0}
\caption{The Young patterns of the $SU(4)$ representations $1$ to
$8$ presented in Tab. IV.} \label{fig:young}
\end{figure}

We summarize some frequently used representations of $SU(4)(SO(6))$
in Tab. IV.
Representations with indices from 1 to 6 are the identity (1d),
the fundamental spinor (4d) and its complex conjugation (4d),
the rank-2 anti-symmetric tensor (6d),
the rank-2 symmetric tensor (10d) and its complex conjugation (10d),
the adjoint (15d) representations, respectively.
On the other hand, the Young pattern is often convenient for
the representations of $SU(N)$ group.
The Young patterns of the representations from 1 to 8 in Tab. IV
are shown in Fig. \ref{fig:young}.


\bibliography{spin32}

\end{document}